\documentclass[twocolumn]{aastex62}
\usepackage[varg]{txfonts}
\usepackage{graphicx}
\usepackage{color}
\begin{document}

\title{JCMT POL-2 and ALMA polarimetric observations of 6000--100 au scales in the protostar B335: linking magnetic field and gas kinematics in observations and MHD simulations}

\author{Hsi-Wei Yen}
\affiliation{Academia Sinica Institute of Astronomy and Astrophysics, P.O. Box 23-141, Taipei 10617, Taiwan}
\affiliation{European Southern Observatory (ESO), Karl-Schwarzschild-Str. 2, D-85748 Garching, Germany}

\author{Bo Zhao}
\affiliation{Max-Planck-Institut f\"ur extraterrestrische Physik (MPE), Garching, Germany, 85748}

\author{I-Ta Hsieh}
\affiliation{Academia Sinica Institute of Astronomy and Astrophysics, P.O. Box 23-141, Taipei 10617, Taiwan}

\author{Patrick Koch}
\affiliation{Academia Sinica Institute of Astronomy and Astrophysics, P.O. Box 23-141, Taipei 10617, Taiwan}

\author{Ruben Krasnopolsky}
\affiliation{Academia Sinica Institute of Astronomy and Astrophysics, P.O. Box 23-141, Taipei 10617, Taiwan}

\author{Chin-Fei Lee}
\affiliation{Academia Sinica Institute of Astronomy and Astrophysics, P.O. Box 23-141, Taipei 10617, Taiwan}

\author{Zhi-Yun Li}
\affiliation{Astronomy Department, University of Virginia, Charlottesville, VA 22904, USA}

\author{Sheng-Yuan Liu}
\affiliation{Academia Sinica Institute of Astronomy and Astrophysics, P.O. Box 23-141, Taipei 10617, Taiwan}

\author{Nagayoshi Ohashi}
\affiliation{Subaru Telescope, National Astronomical Observatory of Japan, 650 North A'ohoku Place, Hilo, HI, 96720, USA}

\author{Shigehisa Takakuwa}
\affiliation{Department of Physics and Astronomy, Graduate School of Science and Engineering, Kagoshima University, 1-21-35 Korimoto, Kagoshima, Kagoshima 890-0065, Japan}
\affiliation{Academia Sinica Institute of Astronomy and Astrophysics, P.O. Box 23-141, Taipei 10617, Taiwan}

\author{Ya-Wen Tang}
\affiliation{Academia Sinica Institute of Astronomy and Astrophysics, P.O. Box 23-141, Taipei 10617, Taiwan}

\correspondingauthor{Hsi-Wei Yen}
\email{hwyen@asiaa.sinica.edu.tw}

\begin{abstract}
We present our analysis of the magnetic field structures from 6000 au to 100 au scales in the Class 0 protostar B335 inferred from our JCMT POL-2 observations and the ALMA archival polarimetric data. 
To interpret the observational results, we perform a series of (non-)ideal MHD simulations of the collapse of a rotating non-turbulent dense core, whose initial conditions are adopted to be the same as observed in B335, and generate synthetic polarization maps. 
The comparison of our JCMT and simulation results suggests that the magnetic field on a 6000 au scale in B335 is pinched and well aligned with the bipolar outflow along the east--west direction. 
Among all our simulations, the ALMA polarimetric results are best explained with weak magnetic field models having an initial mass-to-flux ratio of 9.6.
However, we find that with the weak magnetic field, the rotational velocity on a 100 au scale and the disk size in our simulations are larger than the observational estimates by a factor of several. 
An independent comparison of our simulations and the gas kinematics in B335 observed with the SMA and ALMA favors strong magnetic field models with an initial mass-to-flux ratio smaller than 4.8.
We discuss two possibilities resulting in the different magnetic field strengths inferred from the polarimetric and molecular-line observations, (1) overestimated rotational-to-gravitational energy in B335 and (2) additional contributions in the polarized intensity due to scattering on a 100 au scale.
\end{abstract}


\section{Introduction}
Stars form via gravitational collapse of dense cores \citep{Shu87}, which are magnetized \citep{Crutcher12}.
During the collapse, the magnetic field lines are expected to be dragged inward by collapsing material, 
and the magnetic flux in the inner envelope around a central protostar increases \citep[e.g.,][]{Li96, Galli06}.
As the magnetic field strength increases in the inner envelope, 
the magnetic field can slow down the infalling and rotational motions of the collapsing material more efficiently \citep[e.g.,][]{Allen03}, 
if the field and matter remain well coupled. 
As a result, the collapsing material is expected to infall toward the center at a velocity slower than the free-fall velocity, 
and its angular momentum is transferred outward, leading to suppression of formation and growth of a rotationally-supported disk \citep[e.g.,][]{Mellon08}.
Signs of infalling motion slower than free fall and removal of angular momentum of collapsing material have been observed in several Class 0 and I protostars, such as L1527 \citep{Ohashi14}, B335 \citep{Yen15b}, TMC-1A \citep{Aso15}, L1551 IRS 5 \citep{Chou14}, and HH 111 \citep{Lee16}.
These results suggest that the magnetic field could play an importance role in the dynamics in collapsing dense cores. 
Nevertheless, it remains unclear as to how efficiently the magnetic field affects the star-forming process. 

Polarized thermal dust continuum emission at (sub-)millimeter wavelengths can be adopted to trace magnetic field structures on scales of hundreds to thousands of au in dense cores \citep{Crutcher12}, 
where dust grains are expected to preferentially align their long axis perpendicular to the magnetic field \citep{Lazarian07}.
Thus, the magnetic field orientation can be inferred by rotating the polarization orientation by 90$\degr$.
In addition, 
in young protostellar sources, the sizes of dust grains on these scales are likely smaller than 100 $\mu$m \citep[e.g.,][]{Li17}, 
so that dust scattering unlikely induces any significant polarized intensity \citep{Kataoka15, Yang16, Yang17}. 
With polarimetric observations at (sub-)millimeter wavelengths, 
magnetic field lines being dragged to form an hour-glass morphology by collapse or being wrapped by rotational motion have been seen in protostellar envelopes around several protostars \citep{Girart06, Attard09, Rao09, Hull14, Davidson14, Cox18, Sadavoy18, Lee18, Maury18, Kwon18} as well as in high-mass star-forming regions \citep{Girart09, Qiu13, Qiu14, bistro, Pattle17}. 
These results suggest an interplay between the magnetic field and the gas motions. 
Therefore, linking observational results of magnetic field structures and gas kinematics could shed light on the role of the magnetic field in the dynamics of collapsing cores. 

B335 is an isolated Bok globule with an embedded Class 0 protostar at a distance of 100 pc \citep{Keene80, Keene83, Stutz08, Olofsson09}.
The size of the dense core in B335 observed at millimeter wavelengths is $\sim$0.1 pc \citep{Saito99, Motte01, Shirley02}, 
and the core is slowly rotating \citep{Saito99, Yen11, Kurono13}. 
Infalling and rotational motions on scales from 100 au to 3000 au have been observed in molecular lines with single-dish telescopes and interferometers \citep{Zhou93, Zhou95, Choi95, Evans05, Evans15, Saito99, Yen10, Yen11, Yen15b, Kurono13}.
Nevertheless, no sign of Keplerian rotation was observed with the Atacama Large Millimeter/submillimeter Array (ALMA) at an angular resolution of 0\farcs3 \citep[30 au;][]{Yen15b}, and the envelope rotation on a scale of 100--1000 au in B335 is an order of magnitude slower than in other Class 0 and I protostars surrounded by a Keplerian disk with a size of tens of au \citep{Yen15a}.  
The presence of a small disk less than 10 au and the slow envelope rotation hints at the effects of the magnetic field on the gas kinematics in B335.
In addition, ALMA observations in the C$^{18}$O and H$^{13}$CO$^+$ lines show no detectable difference in the infalling velocities of neutral and ionized gas on a 100 au scale with a constraint on the upper limit of the ambipolar drift velocity of 0.3 km/s, 
suggesting that the magnetic field likely remains well coupled with the matter in the inner envelope in B335 \citep{Yen18}.
The magnetic field structures on a 1000 au scale in B335 also show signs of being dragged toward the center and become pinched, as inferred from the ALMA polarimetric observations \citep{Maury18}.
Therefore, B335 is an excellent target to investigate the interplay between the magnetic field and gas motions and the effects of the magnetic field on the dynamics in collapsing dense cores. 

Theoretical simulations show that the importance of the magnetic field on the dynamics in collapsing dense cores is closely related to the magnetic field strength, coupling between the magnetic field and matter, and alignment between the magnetic field and rotational axis \citep{Mellon08, Mellon09, Li11, Li13, Dapp12, Joos12, Padovani14, Tsukamoto15, Zhao16, Zhao18a, Masson16}. 
In the present work, we study these three physical parameters by linking the information from the polarimetric and molecular-line observations of B335 and by comparing them with theoretical simulations. 
We conducted polarimetric observations at submillimeter wavelength with the James Clerk Maxwell Telescope (JCMT) to map the magnetic field structures on a scale of thousands of au in B335. 
We additionally obtained the ALMA polarimetric data at millimeter wavelength from the public archive in order to study the magnetic field structures on scales from the dense core to the inner envelope in B335. 
The details of the observations and the observational results are presented in Section \ref{ob}.
To analyze the observed magnetic field structures, 
we performed a series of (non-)ideal magnetohydrodynamics (MHD) simulations and generated synthetic polarization maps to compare with the observations.  
The simulation setup and results are described in Section \ref{S.IC}.
In Section \ref{obvssim}, we compare the magnetic field structures from the observations and the simulations. 
With these polarimetric data and our MHD simulations as well as the observational results of the gas kinematics from our previous studies with the SMT, SMA, and ALMA, 
we discuss the magnetic field strength and the coupling between the field and matter in B335 in Section \ref{discuss}.

\section{Polarimetric observations}\label{ob}
\subsection{JCMT POL-2 observations}

\begin{table*}
\caption{JCMT POL-2 detections}\label{pol2data}
\centering
\begin{tabular}{cccccccc}
\hline\hline
RA & Dec & PP & $\Delta$PP & PA & $\Delta$PA & $I_{\rm p}$ & $\Delta I_{\rm p}$ \\
 & & (\%) & (\%) & ($\degr$) & (\degr) & (mJy/beam) & (mJy/beam) \\
\hline\hline
19:37:02.28 & +07:34:01.8 & 3.8 & 1.0 & 18 &  7 &  6.2 & 1.5 \\ 
19:36:59.86 & +07:34:01.8 & 2.2 &  0.7 & $-19$ & 8 & 4.8 & 1.5 \\         
19:37:00.66 & +07:34:13.8 & 0.7 & 0.1 &  5 &   6 &  6.8 & 1.5 \\         
19:36:59.86 & +07:34:13.8 & 2.8 &  0.6 & 9 & 6 &  6.4 &  1.5 \\         
19:36:59.86 & +07:34:25.8 &  4.1 &  1.1 &   $-1$ & 7 &   5.5 &  1.5 \\
19:37:00.66 & +07:34:37.8 &   8.3 &  2.4 &  2 &  8 &    5.2 &   1.5 \\
19:36:59.86 & +07:34:37.8 &  15.3 &  3.5 &  47 &  6 &   6.5 & 1.5 \\
\hline\hline
\end{tabular}
\tablecomments{PP, PA, and $I_{\rm p}$ present the polarization percentage, the position angle of the polarization orientation, and the polarized intensity, respectively. $\Delta$PP, $\Delta$PA, and $\Delta I_{\rm p}$ are their uncertainties.}
\end{table*}

The observations with the JCMT were conducted on April 18 and May 5 and 12 in 2017. 
During the observations, the 225 GHz opacity ranged from 0.03 to 0.08. 
Polarized continuum emission was observed at 850 $\mu$m and 450 $\mu$m simultaneously with the polarimeter POL-2 \citep{Friberg16} and the continuum receiver SCUBA-2 \citep{Holland13}. 
The POL-2 observing mode at 450 $\mu$m is not fully commissioned. 
In this paper, we present the results at 850 $\mu$m.
The angular resolution of JCMT at 850 $\mu$m is 14$\arcsec$.
The total on-source observing time was 6.8 hours. 
The Daisy observing mode was adopted.
With this observing mode, the exposure time of the central region within a radius of 3$\arcmin$ is above 80\% of the total observing time. 

The data were reduced with the software {\it Starlink} \citep{Currie14} and the task {\it pol2map}. 
The data were first reduced with the default pixel size of 4$\arcsec$ and the procedure {\it makemap} to obtain initial Stokes {\it I} maps. 
These initial Stokes {\it I} maps were adopted as the reference to correct the instrumental polarization and to generate masks in the subsequent data reduction process. 
Then, the final Stokes {\it IQU} maps were generated with the procedure {\it skyloop}, 
and the polarized intensity was debiased in this process. 
We binned up the final Stokes {\it IQU} maps from the default pixel size of 4$\arcsec$ to 12$\arcsec$ to improve the signal-to-noise ratio (S/N).
The final pixel size is slightly smaller than the angular resolution of 14$\arcsec$.
The achieved noise level is 1.6 mJy Beam$^{-1}$ in Stokes {\it I} and 1.5 mJy Beam$^{-1}$ in the polarized intensity ($I_{\rm p}$) with the pixel size of 12$\arcsec$.
Our detection criteria of the polarized emission are $I_{\rm p}$ over its uncertainty ($\Delta I_{\rm p}$), $I_{\rm p}/\Delta I_{\rm p} >3$, and the S/N in Stokes {\it I} larger than 5.
This leads to seven detections (Table \ref{pol2data}).
The uncertainties in the polarization orientations, which are $\Delta I_{\rm p}/2I_{\rm p}$ \citep[e.g.,][]{Hull14}, range from 6$\degr$ to 8$\degr$.

\subsection{ALMA observations}
The ALMA polarimetric data at 1.3 mm analyzed here were retrieved from the archive (project code: 2013.1.01380.S).  
The details of the ALMA observations were described in \citet{Maury18}.
We reduced the data with the calibration script provided in the archive using the software Common Astronomy Software Applications \citep[CASA;][]{McMullin07} of version 4.7.0, and we further performed self calibration on the phase.
The images were generated with Briggs weighting with a robust parameter of 0.5.
The achieved angular resolution is 0\farcs67 $\times$ 0\farcs47 ($\sim$70 au $\times$ 50 au). 
The achieved noise level is 50 $\mu$Jy in Stokes {\it I} and 15 $\mu$Jy in both Stokes {\it Q} and {\it U}.
We debiased $I_{\rm p}$ with $I_{\rm p} = \sqrt{Q^2 + U^2 - {\sigma_{Q,U}}^2}$, where $\sigma_{Q,U}$ is the noise level in Stokes {\it Q} and {\it U} \citep{Wardle74,Simmons85}. 
We binned up the Stokes {\it Q} and {\it U} maps to have a pixel size of 0\farcs35, approximately half of the beam size, to extract the polarization detections. 
Our detection criteria of the polarized emission are S/N higher than 3 in Stokes {\it I} and $I_{\rm p}/\Delta I_{\rm p} > 3$. 
The expected uncertainties in the polarization orientations are $\lesssim$9\degr.

\subsection{Observational results}\label{obsec}

\begin{figure}
\centering
\includegraphics[width=8cm]{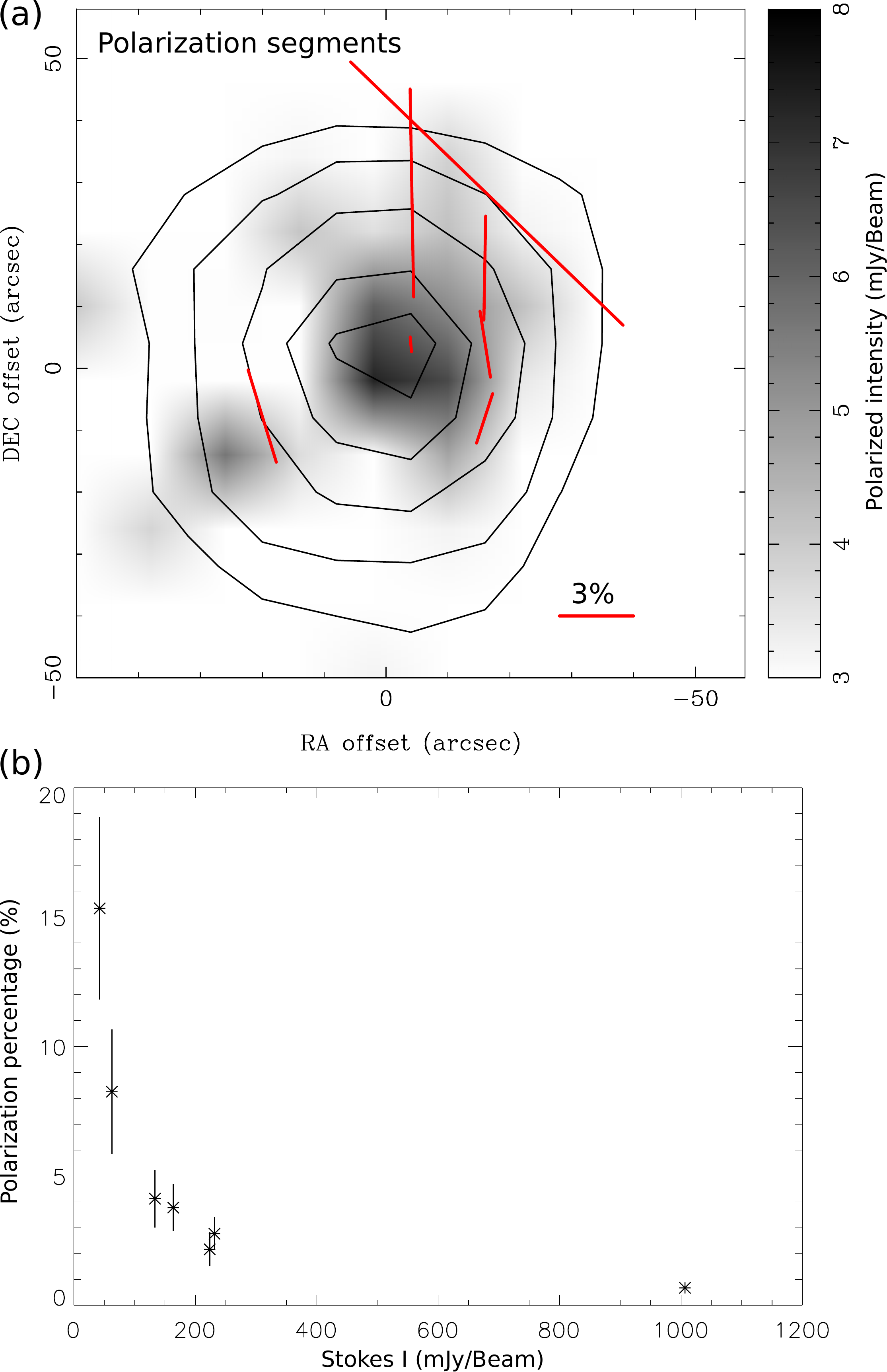}
\caption{JCMT POL-2 results of B335 at 850 $\mu$m. (a) Stokes {\it I} map (contours) overlaid on polarized intensity map (grey scale). Segments show the polarization orientations detected at a level above 3$\sigma$, and their lengths are proportional to the polarization percentage. The center of the map is 19$^{\rm h}$37$^{\rm m}$00\fs 93 +7$^{\rm d}$34$^{\rm m}$09\fs 8. (b) Polarization percentage as a function of Stokes {\it I} intensity.}\label{polmap}
\end{figure}

\begin{figure*}
\centering
\includegraphics[width=16cm]{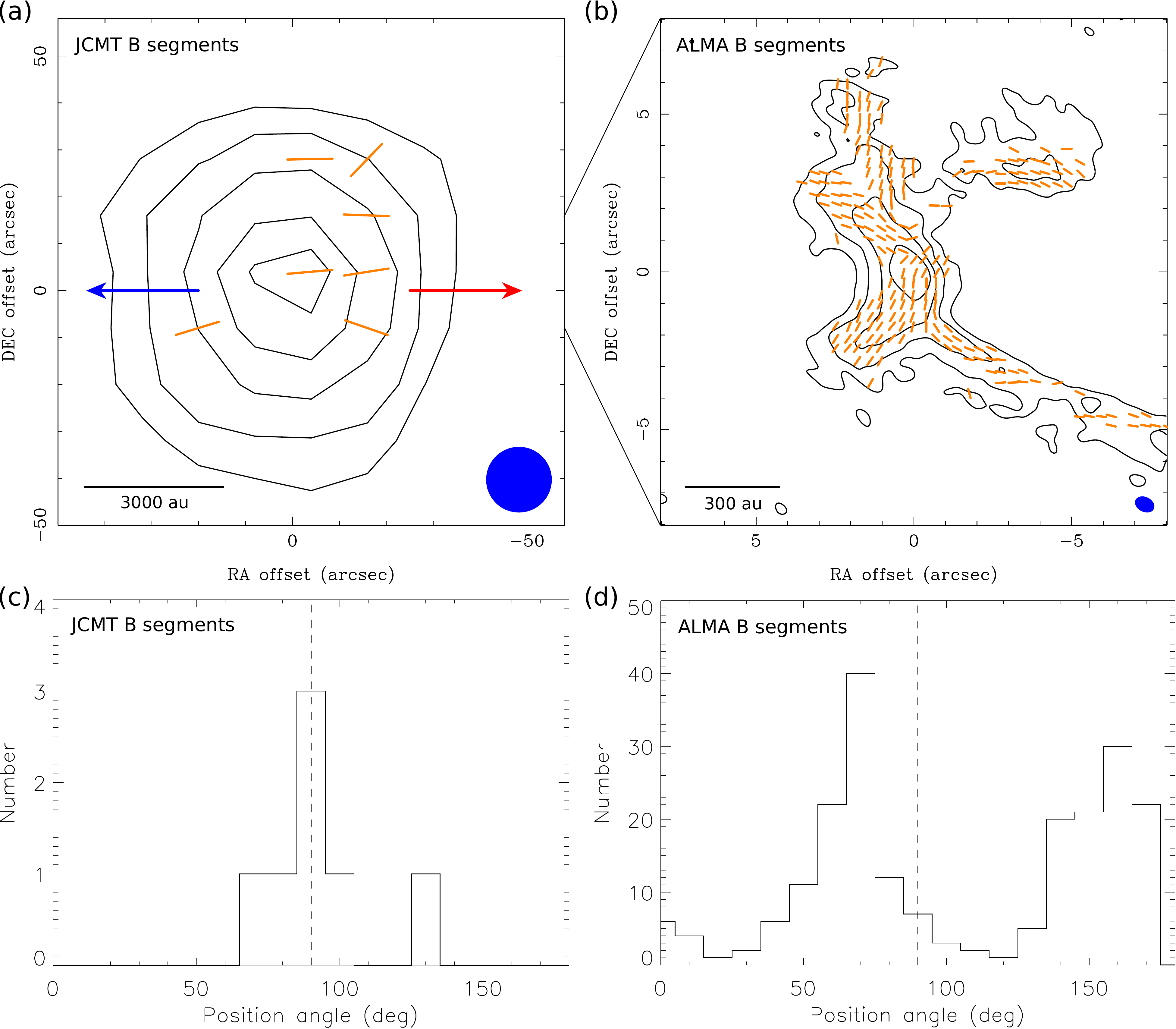}
\caption{(a) \& (b) Magnetic field orientations (orange segments) inferred by rotating the polarization orientations by 90$\degr$ obtained with the JCMT POL-2 and ALMA observations, respectively. Contours shows the Stokes {\it I} maps. Blue and red arrows present the directions of the blue- and redshifted outflows. Blue ellipses at the bottom right corners show the angular resolutions. (c) and (d) Number distributions of the magnetic field orientations. Dashed vertical lines denote the PA of the outflow axis.}\label{bmap}
\end{figure*}

Figure \ref{polmap}a presents the maps of the Stokes {\it I} and polarized intensity at 850 $\mu$m of \object{B335} observed with the JCMT. 
The diameter of the observed dense core is $\sim$8000 au. 
The polarized emission is primarily detected in the western region, and there is one detection in the east. 
We note that there is no detection in the northeast and the south.
In addition, there is only one detection out of four pixels in the central 20$\arcsec$ region.
All the other detections are at outer radii larger than 20$\arcsec$.
Figure \ref{polmap}b shows the polarization percentage as a function of Stokes {\it I} intensity. 
The polarization percentage clearly decreases with increasing Stokes {\it I} intensity, as reported in several dense cores \citep[e.g.,][]{Wolf03}.
The polarization precentage close to the center is 0.7\%. 
This is much lower than the median polarization percentage of 10\% on a scale of 1000 au (10$\arcsec$), which is comparable to the JCMT beam size of 14$\arcsec$, detected with the SMA at an angular resolution of 5$\arcsec$ \citep{Galametz18}. 
As the ALMA observations at an angular resolution of 0\farcs7 show complex structures in the polarized emission \citep[Fig.~\ref{bmap}b and ][]{Maury18}, 
the low polarization percentage close to the center observed with the JCMT is most likely due to canceling of different polarization orientations within the larger JCMT beam.

Figure \ref{bmap}a presents the orientations of the magnetic field segments inferred by rotating the polarization orientations by 90\degr.
The inferred magnetic field orientations are organized and are along the east--west direction, except for one segment in the northwest with an orientation clearly deviated from the others.  
Figure \ref{bmap}c presents the number distribution of the inferred magnetic field orientations from the JCMT polarization detections. 
The mean orientation of the magnetic field segments is 99$\degr$ with a standard deviation of 20$\degr$. 
Excluding the segment in the northwest with a position angle (PA) of 137$\degr$, the mean orientation becomes 92$\degr$ with a standard deviation of 12$\degr$, comparable to the 2$\sigma$ uncertainty in the polarization angle.
The mean orientation of the magnetic field is consistent with the direction of the outflow having a PA of 90$\degr$--99$\degr$ observed in the CO lines in B335 \citep{Hir88, Yen10, Hull14}.

We note that the inferred magnetic field orientations from our POL-2 observations are different from the observational results obtained with SCUPOL \citep{Wolf03, Matthews09}.
In the SCUPOL results, the magnetic field orientations have a mean PA  of 3$\degr$ and a standard deviation of 36$\degr$, and are perpendicular to the direction of the outflow \citep{Wolf03}. 
Only two SCUPOL detections in the northwest show orientations along the east--west direction, similar to our POL-2 detections \citep{Wolf03, Matthews09}.
In addition, the polarization percentage observed with SCUPOL is a factor of two to three higher than that in our POL-2 observations \citep{Wolf03}. 
A detailed comparison between the performance of POL-2 and SCUPOL has been presented in \citet{bistro} with the observations of \object{OMC~1}. 
The comparison shows that the observed polarization orientations and percentages with POL-2 and SCUPOL are consistent in the central bright regions of \object{OMC~1}. 
On the other hand, 
in the outer faint regions, the SCUPOL detections tend to show higher polarization percentages, 
and the difference in the polarization orientations can be as large as 30$\degr$ to 50$\degr$ \citep[Fig.~7 and 8 in][]{bistro}.
This discrepancy is possibly due to the low S/N of the SCUPOL data in the faint region \citep{bistro}. 
As the noise level of our POL-2 observations is a factor of three lower than that of the SCUPOL data \citep{Matthews09}, 
the difference between our POL-2 and the SCUPOL results can be attributed to the higher noise level in the SCUPOL data. 

Figure \ref{bmap}b presents the Stokes {\it I} map of the 1.3 mm continuum emission on a 1000 au scale in \object{B335} overlaid with the magnetic field orientations from the ALMA polarimetric observations.
The details of the ALMA results have been described by \citet{Maury18}.
The polarization detections obtained with ALMA are primarily along the wall of the outflow cavity, 
and there is an additional patch of detections in the north. 
The area of the polarized emission detected with ALMA is within the JCMT beam at the intensity peak position. 
The number distribution of the magnetic field orientations inferred from the ALMA polarization detections is shown in Fig.~\ref{bmap}d. 
The distribution has a double peak at a PA of 70$\degr$ and 160$\degr$, as shown in \citet{Maury18}.
This number distribution with the double peaks is consistent with the expectation from a pinched magnetic field along the east--west direction \citep{Li96, Galli06}, 
as discussed and demonstrated in \citet{Maury18} with the results from their non-ideal MHD simulations. 

In summary, our JCMT results and the ALMA archival data show that the magnetic field on scales from 1000 au to 6000 au in B335 is most likely along the east--west direction, which is aligned with the outflow direction within 10$\degr$ on the plane of the sky, 
and the magnetic field is likely pinched on a 1000 au scale in B335. 
We note that the SMA polarimetric observations at an angular resolution of $\sim$5$\arcsec$ show orangized magnetic field orientations  tilted from the outflow axis by 35$\degr$ on a 1000 au scale in B335 \citep{Galametz18}. 
This could be due to the limited angular resolution and sensitivity of the SMA that likely only detected a part of the pinched magnetic field, leading to the mean orientation misaligned with the outflow axis.  
 
\section{MHD simulations}\label{S.IC}

To study the strength of the magnetic field and its effect on the gas kinematics in B335, 
we carried out three-dimensional (3D) non-ideal MHD simulations using ZeusTW code \citep{Krasnopolsky10} to model the collapse of a magnetized non-turbulent dense core, and generated synthetic polarization maps from the simulations to compare with the observations.

\subsection{Simulation setup}
In our simulations, among the three non-ideal MHD effects, we only consider ambipolar diffusion (AD) because it is the most efficient mechanism in magnetic field decoupling on a scale from 100 au to thousands of au in a collapsing dense core \citep{Zhao16,Zhao18a}. 
The magnetic diffusivity of ambipolar diffusion is self-consistently computed at run-time using the equilibrium chemical network from \citet{Zhao16}, 
and is closely related to size of dust grains and cosmic-ray ionization \citep{Zhao16,Zhao18b}.
For comparison, we also performed ideal MHD simulations, which are independent on grain size and cosmic-ray ionization. 

We prescribe the initial conditions of the dense core in our simulations to approximately match the observed properties of B335. 
In B335, the mass of the dense core within a radius of 6000 au is estimated to be 0.5--1.8 $M_\sun$ based on the single-dish observational results\footnote{We corrected the values in the literature for the distance from 250 pc to 100 pc \citep{Stutz08, Olofsson09}.} in the millimeter continuum emission and in the C$^{18}$O and H$^{13}$CO$^+$ lines \citep{Saito99, Harvey03, Evans05, Kurono13}. 
Thus, in our simulations, the initial core was set to have a total mass of 1.0~$M_{\sun}$ and an outer radius of $10^{17}$~cm (6684~au), 
and its density distribution is uniform and spherical. 
We note that the outer radius of the initial core in our simulations is smaller than the core radius of $\sim$10\,000--14\,000 au in B335 \citep{Saito99}.
Nevertheless, the adopted outer radius of the initial core is a factor of two larger than the estimated radius of the dynamically infalling region in B335 \citep{Choi95, Harvey03, Evans05, Evans15}, where the interplay between the magnetic field and the gas kinematics becomes prominent. 
The initial core is rotating as a solid body in our simulations.  
The angular speed of its solid-body rotation is adopted to be $4\times10^{-14}$~s$^{-1}$, which is the same as the core rotation in B335 measured from the velocity gradient in the C$^{18}$O and H$^{13}$CO$^+$ emission observed with single-dish telescopes \citep{Saito99, Yen11, Kurono13}. 
The corresponding ratio of rotational-to-gravitational energy $\beta_{\rm rot}$ is around 0.4\%. 

In our simulations, the magnetic field uniformly threads the initial core along the rotation axis. 
Thus, our simulations are initially axisymmetric. 
We focus on axisymmetric models because the misalignment between the magnetic field and the rotational axis is observed to be small on the plane of the sky ($<$10$\degr$; Section~\ref{obsec}). 
Three different magnetic field strengths $B_0$ of 42.5~$\mu$G, $21.3~\mu$G, and $10.2~\mu$G were adopted for strong, intermediate, and weak field cases. 
This gives a dimensionless mass-to-flux ratio $\lambda$ ($\equiv{M_{\rm c} \over {\pi R_{\rm c}^2 B_0}}2\pi\sqrt{G}$) 
of 2.4, 4.8, and 9.6, respectively.
The adopted magnetic field strength is the same as the observational estimate of 10--40 $\mu$G on a 0.1 pc scale in B335 from the infrared polarimetric observations using the \citet{Chandrasekhar53} method by \citet{Bertrang14}.
After having narrowed down the main physical quantities (see Section~\ref{obvssim}), we also carried out two additional simulations with a slight misalignment with an angle of 15$\degr$ between the magnetic field and the rotation axis.

In addition, we varied the minimum grain size $a_{\rm min}$ and the cosmic-ray ionization rate $\zeta_0$ because they play an important role in the magnetic field decoupling in the collapsing envelope, 
which in turn determines the field strength and the degree of pinch during the evolution as well as the disk size formed eventually. 
In our simulations, the slope of the grain size distribution and the maximum grain size $a_{\rm max}$ are fixed at $-3.5$ and 0.25~$\mu$m, respectively, 
and we adopted two different values of $a_{\rm min}$, 0.005~$\mu$m and 0.1~$\mu$m. 
Those models with $a_{\rm min}$ = 0.005~$\mu$m and 0.1~$\mu$m are denoted as MRN and tr-MRN \citep[Mathis-Rumpl-Nordsieck;][]{Mathis77}, respectively. 
As shown by \citet{Zhao16}, the tr-MRN case strongly promotes disk formation due to the enhanced AD in the collapsing envelope. 
The simulations without the non-ideal MHD effect of AD are denoted as ideal.
The models with a typical $\zeta_0$ of 10$^{-17}$ s$^{-1}$ are denoted as CR1, and those with a higher $\zeta_0$ of $5 \times 10^{-17}$ s$^{-1}$ as CR5.
The numbers in the names of our models present the initial mass-to-flux ratios $\lambda$ in those models.
All our simulation models are summarized in Table~\ref{mhdtab}. 
The results of our non-ideal MHD simulations were extracted when the total mass of the central protostar and (if any) rotationally supported disk ($M_{\rm \star+disk}$) reached 0.04 $M_\sun$ and 0.1 $M_\sun$, which is in the range of the observational estimate of 0.04--0.2 $M_\sun$ \citep{Yen15b, Evans15}. 
The results of the ideal MHD simulations were extracted when $M_{\rm \star+disk}$ was 0.02 $M_\sun$ because these simulations fail to evolve longer due to the build-up and growth of MHD instabilities close to the center \citep{Zhao11}.

\subsection{Simulation results}
\label{S.Result}

\begin{table}
\caption{Summary of MHD simulations}\label{mhdtab}
\centering
\begin{tabular}{cccccc}
\hline\hline
Model & $\lambda$ & $\zeta$ & $M_{\rm \star+disk}$ & Disk radius \\
 & & (10$^{17}$s$^{-1}$) & ($M_\odot$) & (AU) \\
\hline
 2.4Ideal & 2.4 & 1 & 0.02 & --  \\ 
 4.8Ideal & 4.8 & 1 & 0.02 & --  \\ 
 9.6Ideal & 9.6 & 1 & 0.02 & --  \\ 
 2.4MRN-CR1 & 2.4 & 1 & 0.04 & --  \\ 
 4.8MRN-CR1 & 4.8 & 1 & 0.04 & --  \\ 
 9.6MRN-CR1 & 9.6 & 1 & 0.04 & $\lesssim$10~au $\rightarrow$ 0  \\ 
 9.6MRN-CR5 & 9.6 & 5 & 0.04 & --  \\ 
 2.4tr-MRN-CR1 & 2.4 & 1 & 0.04 & $<$5~au  \\ 
 4.8tr-MRN-CR1 & 4.8 & 1 & 0.04 & $\lesssim$5~au  \\ 
              & 4.8 & 1 & 0.1 & $<$15~au  \\ 
 4.8tr-MRN-CR5 & 4.8 & 5 & 0.04 & $<$5~au  \\ 
              & 4.8 & 5 & 0.1 & --  \\ 
 9.6tr-MRN-CR1 & 9.6 & 1 & 0.04 & 10--20~au  \\ 
              & 9.6 & 1 & 0.1 & 20--50~au \\ 
 9.6tr-MRN-CR5 & 9.6 & 5 & 0.04 & $\lesssim$10~au \\ 
              & 9.6 & 5 & 0.1 & $\lesssim$10~au \\ 
 9.6tr-MRN-CR10 & 9.6 & 10 & 0.1 & $<$5~au $\rightarrow$ 0  \\ 
\hline
 mis-tr-MRN-CR5 & 9.6 & 5 & 0.04 &  10~au \\
               & 9.6 & 5 & 0.1 &  20~au \\
\hline
\hline
\end{tabular}
\tablecomments{Ideal MHD simulations and non-ideal MHD simulations with the minimum grain sizes of 0.005 $\mu$m and 0.1 $\mu$m are denoted as ideal, MRN, and tr-MRN, respectively. 
$\lambda$ is the initial mass-to-flux ratio. 
$\zeta$ is the cosmic-ray ionziation rate. 
$M_{\rm \star+disk}$ is the total mass of the central star+disk system when we extracted the simulation data, and the disk radius is the radius of the rotationally supported disk (if forms) at that time. 
Arrows indicate transient disks.
Model mis-tr-MRN-CR5 is the simulation with the misaligned magnetic field and rotational axis by 15$\degr$.
In the models with the misalignment, 
the central disks are additionally surrounded by spiral structures which extend to tens of au \citep[e.g.,][]{Li13}.}
\end{table}

\subsubsection{Disk formation}\label{disk}

\begin{figure*}
\centering
\includegraphics[width=\textwidth]{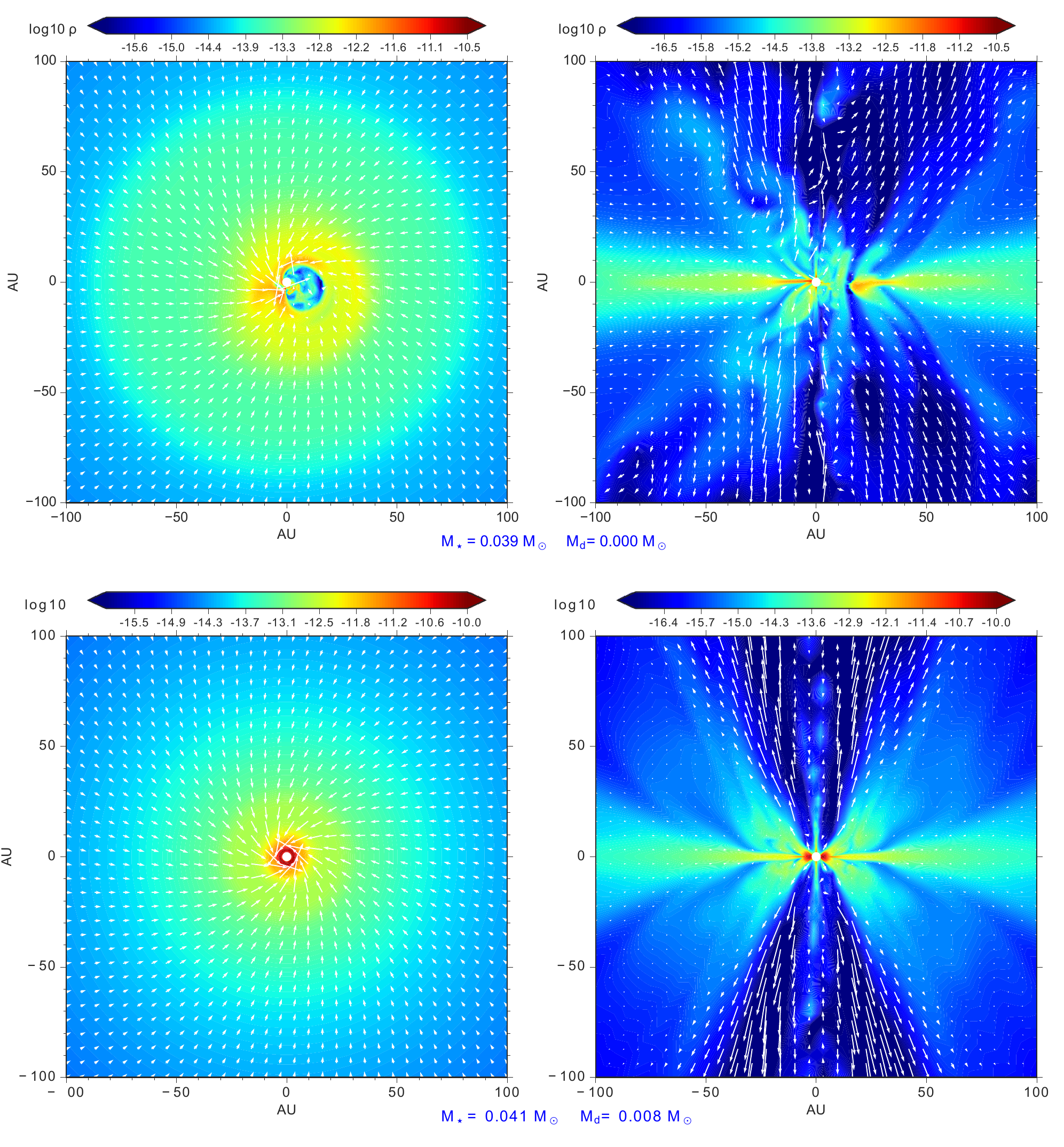}
\caption{Comparison of the disk and outflow morphologies between model 4.8MRN-CR1 (upper panels) and 4.8tr-MRN-CR1 (lower panels), along the equatorial (left) and meridian (right) planes, respectively. Color and arrows present the density and velocity in the simulations, respectively. The masses of the central protostar ($M_\star$) and the disk ($M_{\rm d}$) are labeled below the panels. The disk and outflow structures are well-defined in 4.8tr-MRN-CR1 but  not in 4.8MRN-CR1. }
\label{comp_morph}
\end{figure*}

We found that in the adopted range of $\lambda$, persistent disks only form in the tr-MRN ($a_{\rm min}=0.1~\mu$m) cases,
as discussed in \citet{Zhao16, Zhao18a}. 
The ideal MHD models all fail to evolve to later stages after the formation of the protostar, due to the build-up and growth of MHD instabilities near the central stellar sink \citep{Zhao11}. 
Thus, no disk is formed. 
The disk formation is also hindered in the MRN models. 
Among all the MRN models, 
only the one with the weak field and the low ionization, 9.6MRN-CR1, forms a transient $\sim$10 au disk that shrinks over time due to influx of gas with a low specific angular momentum \citep[see also][]{Zhao18a}.
Rotationally-supported disks generally form in our tr-MRN models.
In these models, 
small dust grains with sizes of tens to hundreds of \AA, which are well coupled to the magnetic field and also exert strong drag to neutral gas, 
are removed. 
As a result, the ambipolar diffusivity is enhanced by one to two orders of magnitude \citep{Zhao16}.
With the enhanced AD, 
magnetic braking becomes less efficient, leading to the formation of a persistent disk. 
The radius of the disk typically remains below $\sim$10~au for most tr-MRN models due to the slow core rotation, 
except for model 9.6tr-MRN-CR1 where both magnetic field strength and cosmic-ray ionization rate are low. 
In Fig.~\ref{comp_morph}, we compare the typical MRN and tr-MRN models in terms of disk and outflow morphologies. 
In the 4.8tr-MRN-CR1 case (bottom panels), a small rotationally supported disk with a radius of few au forms and drives a clear bipolar outflow. 
In contrast, in the 4.8MRN-CR1 case, no disk is formed, and there is also no well-defined outflow.

\subsubsection{Pinched magnetic field}\label{pinch}

\begin{figure*}
\centering
\includegraphics[width=\textwidth]{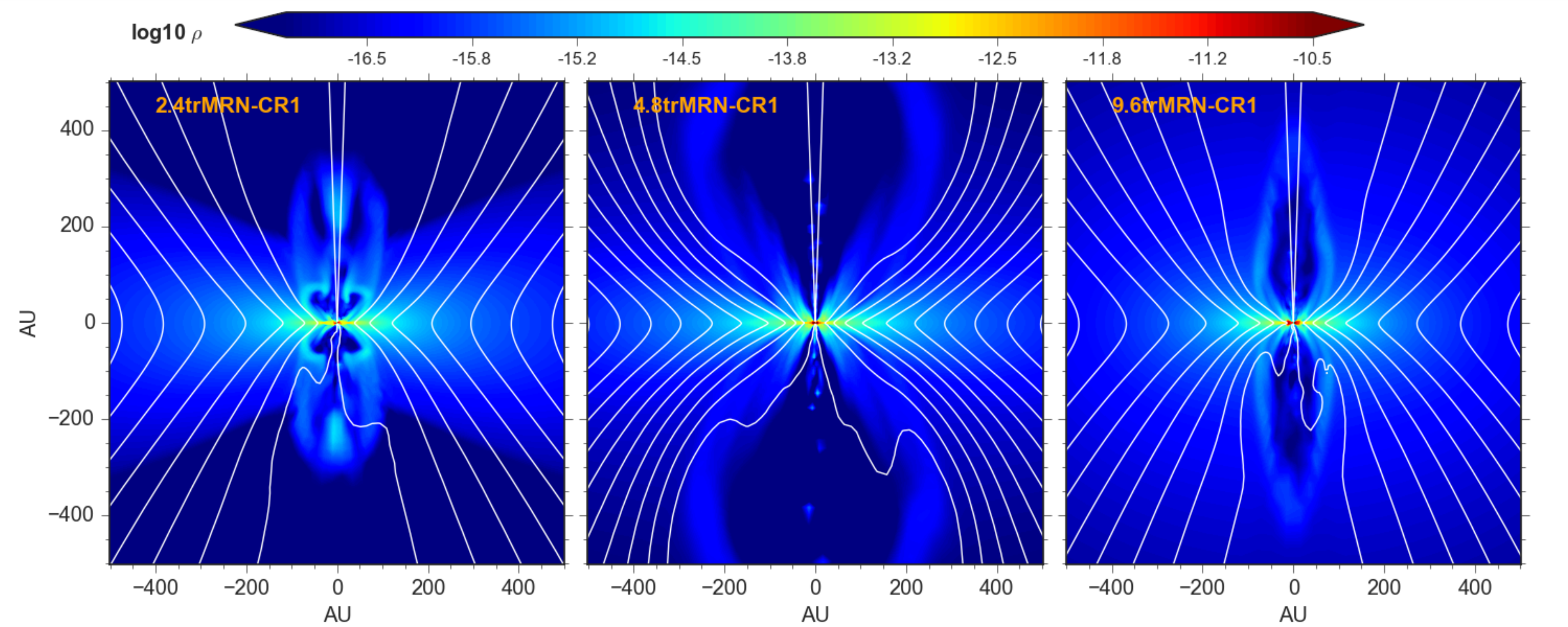}
\includegraphics[width=\textwidth]{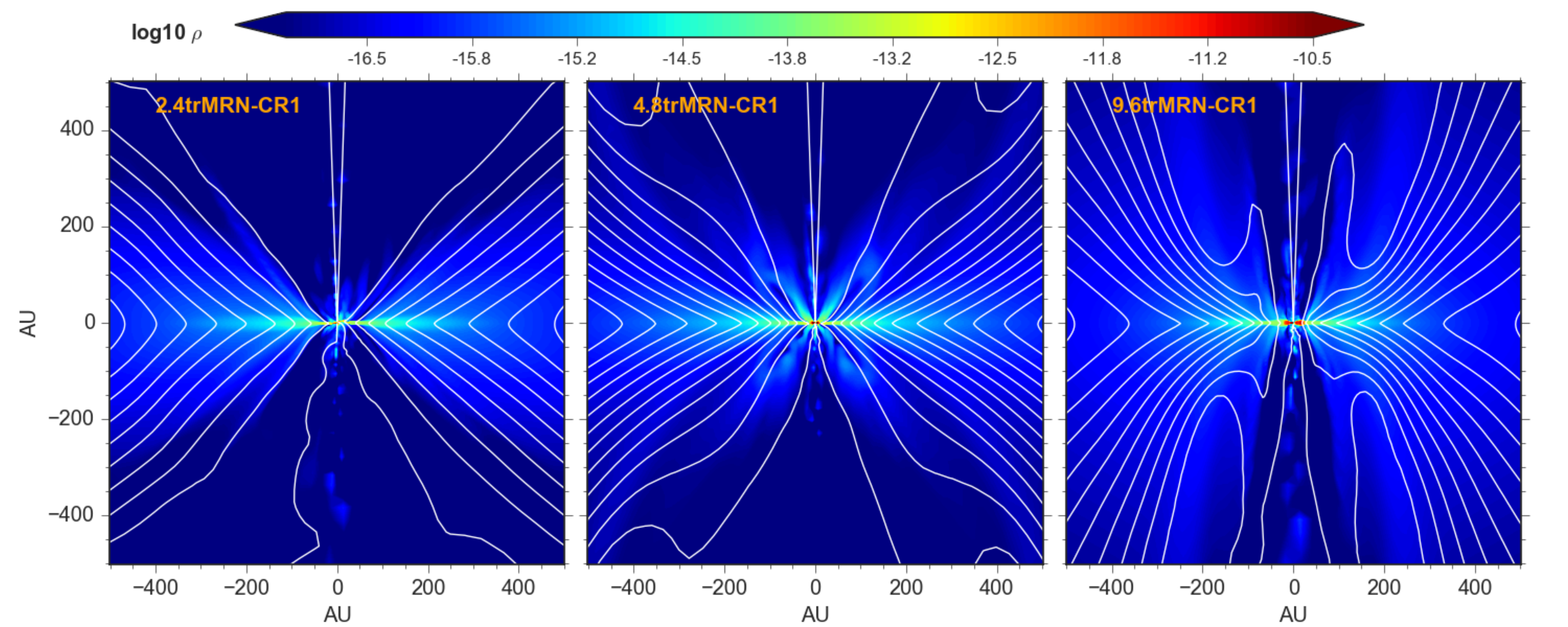}
\caption{Comparison of the structures of the magnetic field lines along the meridian plane (white lines) between cases with different field strengths and at different evolutionary times. The model name is labeled at the upper left corner in each panel. Upper and lower panels present the models at the earlier and later evolutionary times when the star+disk masses are 0.04 $M_\sun$ and 0.1 $M_\sun$, respectively. Color represents the density distributions.}
\label{comp_mhds}
\end{figure*}

In our MHD simulations, the magnetic field morphology generally follows an hour-glass configuration.  
We find that the degree of the pinch of the magnetic field lines is related to evolutionary time, the magnetic field strength in the initial core, and the level of AD. 
Figure~\ref{comp_mhds} presents the magnetic field lines in the tr-MRN models as an example to demonstrate the dependance of the degree of pinch on the initial magnetic field strength and evolutionary time.

\begin{figure*}
\centering
\includegraphics[width=\textwidth]{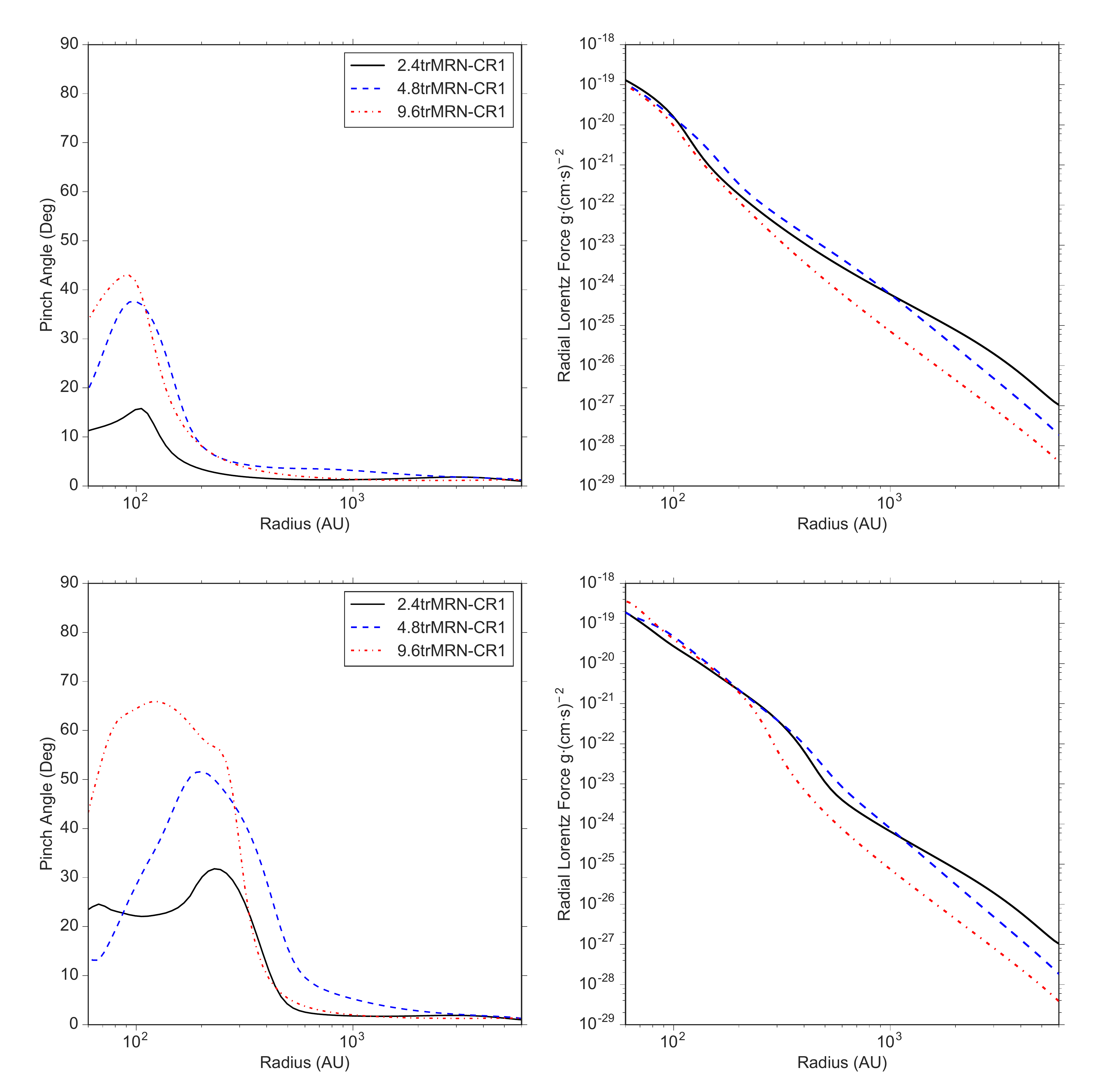}
\caption{Profiles of the pinch angle (left) and radial Lorentz force (right) across the equatorial plane in the models with different magnetic field strengths and at different evolutionary times. A larger pinch angle denotes that the magnetic field lines pinch more severely. Black solid, blue dashed, and red dashed-dotted lines present model 2.4tr-MRN-CR1, 4.8tr-MRN-CR1, and 9.6tr-MRN-CR1, respectively. Upper and lower panels present the models at the earlier and later evolutionary times when the star+disk masses are 0.04 $M_\sun$ and 0.1 $M_\sun$, respectively.}
\label{comp_angle}
\end{figure*}

As shown in Fig.~\ref{comp_mhds}, 
the magnetic field lines are more pinched at the later time with $M_{\rm \star+disk}$ of 0.1 $M_\sun$ than at the early time with $M_{\rm \star+disk}$ of 0.04 $M_\sun$. 
This is because the magnetic field lines are dragged more severely as more material is collapsing into the center. 
The trend is more clearly revealed in the radial profiles of the pinch angles shown in Fig.~\ref{comp_angle}. 
The pinch angle is defined as the angle between the vertical axis and the direction of the magnetic field just above and below the midplane, 
and it is azimuthally averaged.
Thus, the pinch angle here presents the degree of the magnetic field being bent from the original orientation, which is perpendicular to the midplane. 
A pinch angle closer to 90$\degr$ means that the magnetic field is more pinched and become more parallel to the midplane. 
Fig.~\ref{comp_angle} shows that the magnetic field becomes pinched at the inner radii of a few hundred au because of the collapse, 
and the pinch angle at these inner radii increases by 10$\degr$--20$\degr$ from the earlier to later times, 
meaning that the magnetic field is more pinched at the later time.

In addition, the comparison of the pinch angles in the three models in Fig.~\ref{comp_angle} shows that the magnetic field lines pinch more severely in the models with the weaker initial field than the stronger initial field.
We find that the pinch angle is closely regulated by the Lorentz force.
The radial profiles of the Lorentz force are also presented in Fig.~\ref{comp_angle}. 
The Lorentz force along the midplane (pseudo-disk) and pointing away from the center is proportional to $B_z {\delta B_r \over \delta z}$ in the cylindrical coordinate, 
where $B_z$ is the poloidal component and $B_r$ is the radial component. 
In our simulations, the Lorentz force increases with decreasing radii, as the magnetic field lines are dragged inward by the collapse. 
Eventually, the Lorentz force becomes approximately balanced with the gravity on a 100 au scale such that the AD drift velocity is high and the velocity of the magnetic field is low.
The simulation results shown in Fig.~\ref{comp_angle} were extracted at the same evolutionary time, when $M_{\rm \star+disk}$ are 0.04 $M_\sun$ (upper panels) or 0.1 $M_\sun$ (lower panels). 
In these models at the same evolutionary time, their gravitational forces on a 100 au scale are comparable.
Consequently, the Lorentz forces on a 100 au scale in these models with different magnetic field strengths all reach approximately the same magnitude (Fig.~\ref{comp_angle}).
Given the same Loreztz force, 
in the models with the weaker initial magnetic field and thus lower $B_z$, 
the magnetic field lines pinch more severely to have a larger $\delta B_r/\delta z$ along the midplane (pseudo-disk), 
resulting in a larger pinch angle. 

\begin{figure*}
\centering
\includegraphics[width=17cm]{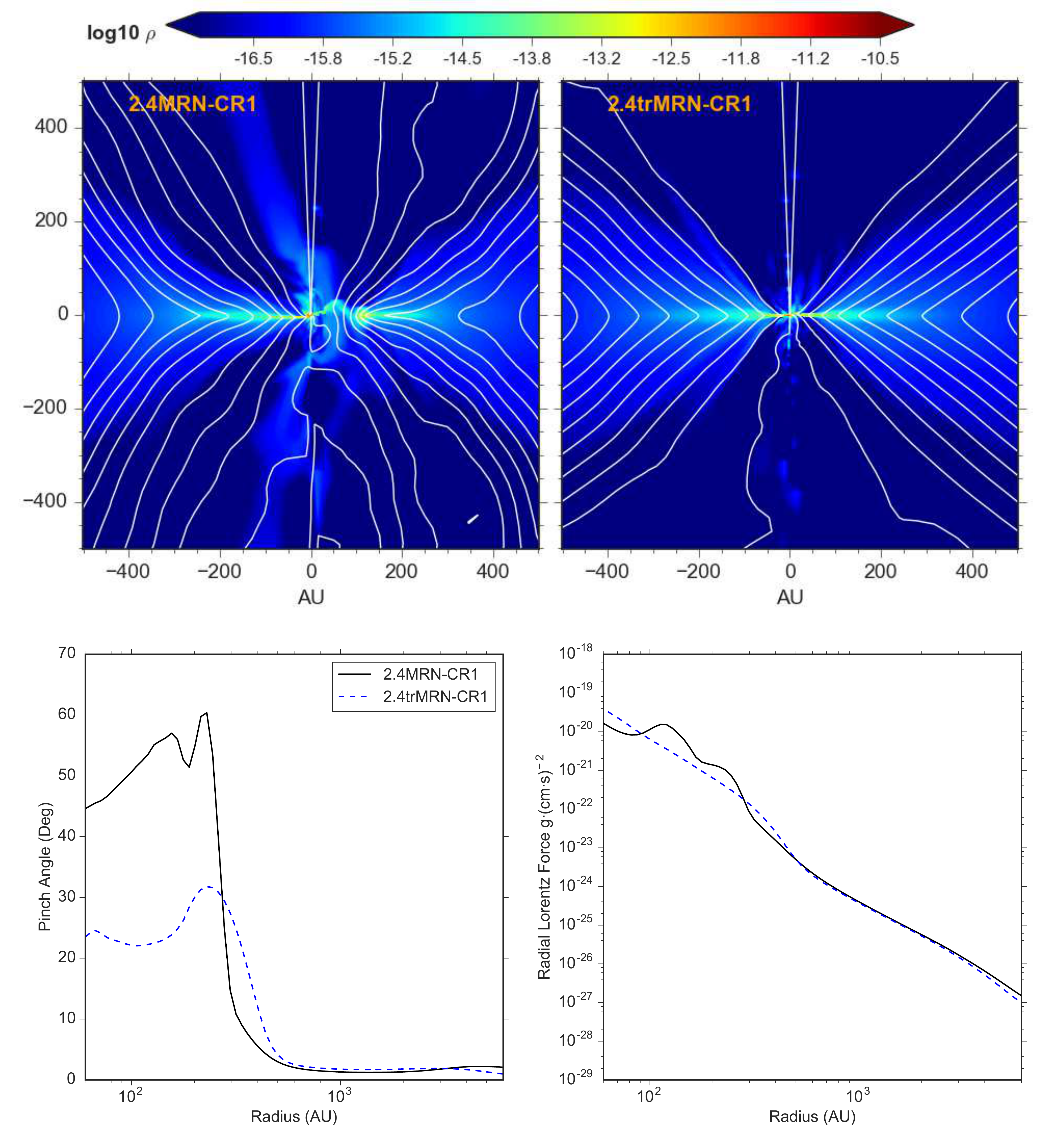}
\caption{Dependence of the pinch of the magnetic field on ambipolar diffusion. 
Upper panels: comparsion of the structures of the magnetic field lines along the 
meridian plane (white lines) between the MRN (left) and tr-MRN (right) models with the same initial $\lambda$ of 2.4 when the central 
star+disk mass is 0.1~$M_\sun$. Color scales present the density distributions. 
Lower panels: profiles of the pinch angle (left) and radial Lorentz force (right) across the 
equatorial plane. A larger pinch angle denotes that the magnetic field lines pinch more severely.}
\label{comp_AD}
\end{figure*}

In Fig.~\ref{comp_AD}, we compare the 2.4MRN-CR1 model with less efficient 
AD and the 2.4trMRN-CR1 model with enhanced AD to study the dependence of the pinch angle on magnetic diffusion, which is in the form of AD in our simulations. 
The magnetic field lines in the MRN model show a sharp kink at a radius of $\sim$200~au, where the pinch angle is as large as 50\degr--60\degr. 
Compared to the MRN model, 
the tr-MRN model shows less pinched field lines with a smaller pinch angle of 20\degr--30$\degr$ at the same radius.
Because of the enhanced AD, magnetic diffusion is more efficient in the tr-MRN model than in the MRN model.
Thus, the magnetic fields in the tr-MRN model are dragged toward the center less efficiently by the collapsing material, which quenches the increase of the Lorentz force, compared to the MRN model. 
As a result, the MRN model has a stronger Lorentz force on an inner scale of 100--300 au than the tr-MRN model (Fig.~\ref{comp_AD}). 
For a given strength of the poloidal component $B_z$, 
the magnetic field is more pinched when the Lorentz force is stronger because of a larger $\delta B_r/\delta z$. 
In these two models, the field strengths of the initial cores are both $\lambda=2.4$. 
Therefore, the magnetic field lines are more pinched in the MRN model with less efficient magnetic diffusion than in the tr-MRN model.

In summary, our MHD simulations suggest that the magnetic field lines pinch more severely in a collapsing dense core with weaker magnetic field, longer time evolution, and lower magnetic diffusion (or AD in our simulations).

\section{Comparison between simulations and observations}\label{obvssim}
To compare the magnetic field structures inferred from the polarimetric observations with the simulations, 
we generated synthetic Stokes {\it IQU} maps from our MHD simultions using the radiative transfer code, Simulation Package for Astronomical Radiative Xfer (SPARX; \url{https://sparx.tiara.sinica.edu.tw/}).
SPARX adopts the properties of dust grains and formulization in \citet{Lee85} to compute Stokes parameters and generate synthetic images. 
In the present paper, we only compare the polarization orientations from the observations and the synthetic images, 
and we do not compare the polarization percentage. 
To properly compute polarization percentage in our models requires more sophisticated calculations of mechanisms and efficiency of grain alignments \citep[e.g.,][]{Lazarian07}, which is beyond the scope of the present paper.
Thus, in our calculations of radiative transfer, the polarization efficiency is assumed to be constant, and we do not consider polarization due to dust scattering \citep[e.g.,][]{Kataoka15, Yang16, Yang17}.

For each model, we generated Stokes {\it IQU} maps at two wavelengths, 850 $\mu$m and 1.3 mm, the same as the observations using SPARX. 
We convolved the model Stokes {\it IQU} maps at 850 $\mu$m and 1.3 mm with the beams of the JCMT and ALMA observations, respectively.
We have also performed ALMA imaging simulations on our models with the most and least pinched magnetic field, 
and found that there is no significant difference between the polarization orientations extracted from the simulated and convolved model maps in the regions where there are ALMA polarization detections  (Appendix \ref{simpol}).
Therefore, we adopt the convolved model maps for the comparison between the models and the observations.
We re-gridded the convolved model maps to have the same image and pixel sizes as our observational maps. 
When we compared the number distributions of the magnetic field orientations between the observations and models, the model field orientations were extracted from the model maps at the same positions as the detections in the observational maps. 
Thus, the total numbers of the segments are the same in the comparison. 

\subsection{Magnetic field on a 6000 au scale}\label{bfield1}

\begin{figure}
\centering
\includegraphics[width=8cm]{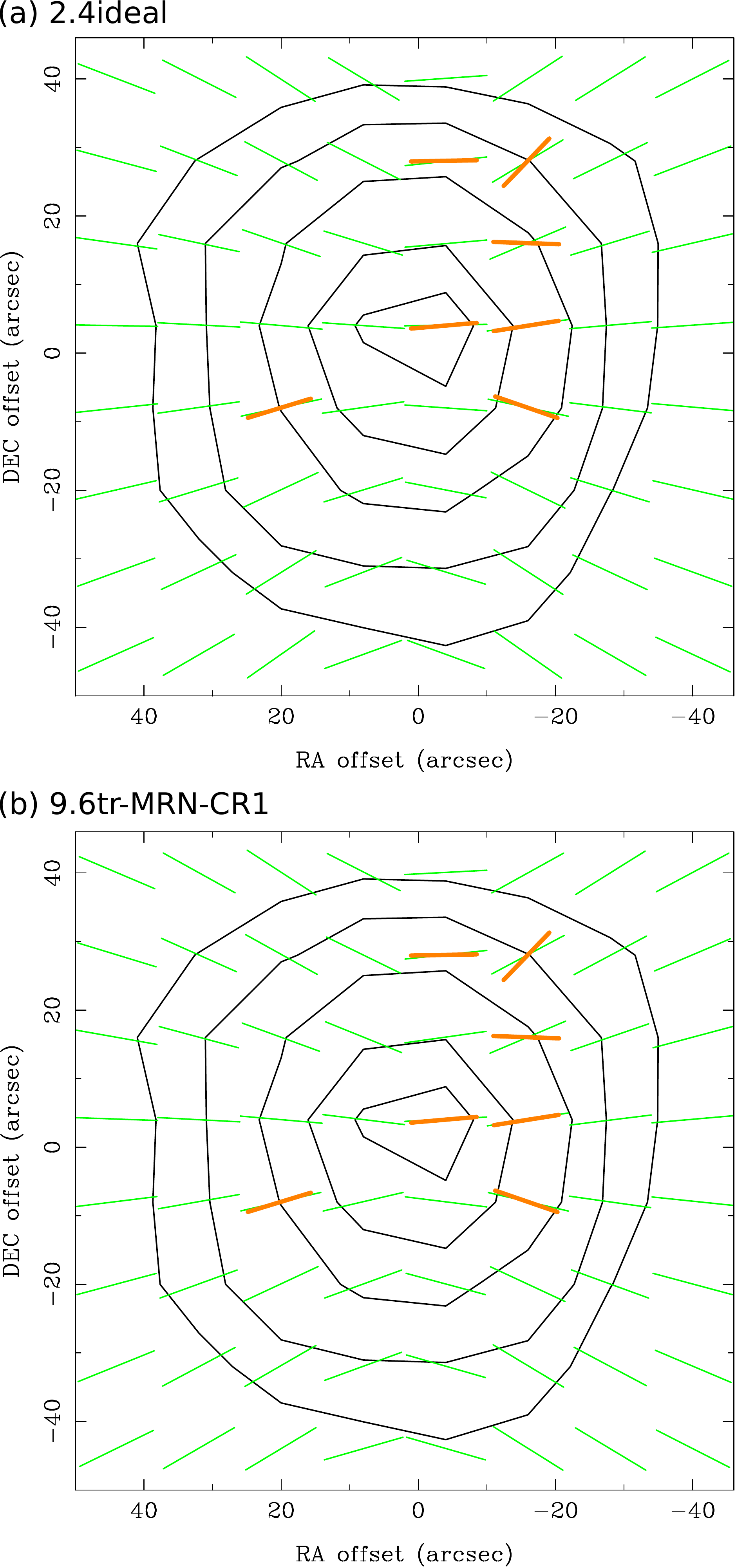}
\caption{Comparison of the magnetic field orientations from the POL-2 observations (orange segments) and from the synthetic maps (green segments) of our models, (a) 2.4Ideal and (b) 9.6tr-MRN-CR1. There is no detectable difference between the two synthetic maps with the JCMT resolution.}\label{simpol1}
\end{figure}

Figure~\ref{simpol1} presents two examples of the comparison between the magnetic field orientations from the JCMT observations and from the models. 
Two extreme cases are shown here, 2.4Ideal having the strongest magnetic field and coupling between the field and the matter, and 9.6tr-MRN-CR1 having the weakest field and coupling, among our models.
The PA of the magnetic field axis in the synthetic maps is adopted to be 90$\degr$.
Figure \ref{simpol1} shows that the magnetic field structures on a 6000 au scale in B335 observed with the JCMT can be well described with the pinched magnetic field in both simulations. 
Although the degrees of the pinch are different among our MHD simulations, 
there is no detectable difference on a 6000 au scale with the angular resolution of the JCMT observations. 
Thus, all our MHD simulations can well explain the magnetic field structures observed with the JCMT, as demonstrated in Fig.~\ref{simpol1}. 
We note that the magnetic field orientation detected at the offset of ($-16\arcsec$, 16$\arcsec$) has a relatively large deviation of 20$\degr$--25$\degr$ from all the models. 
For the other detections, the mean difference between the observed orientations and the models is 5$\degr$--9$\degr$, comparable to the observational uncertainty of 6$\degr$--8$\degr$, 
when the PA of the magnetic field axis in our synthetic maps is adopted to be 87$\degr$--94$\degr$.
Therefore, the comparison between the JCMT observations and our MHD simulations suggest that 
in B335, the magnetic field on a 6000 au scale is pinched and well aligned with the direction of the bipolar outflow within a few degrees on the plane of the sky. 

\subsection{Magnetic field on a 1000 au scale}\label{smallB}

\begin{figure*}
\centering
\includegraphics[width=15.3cm]{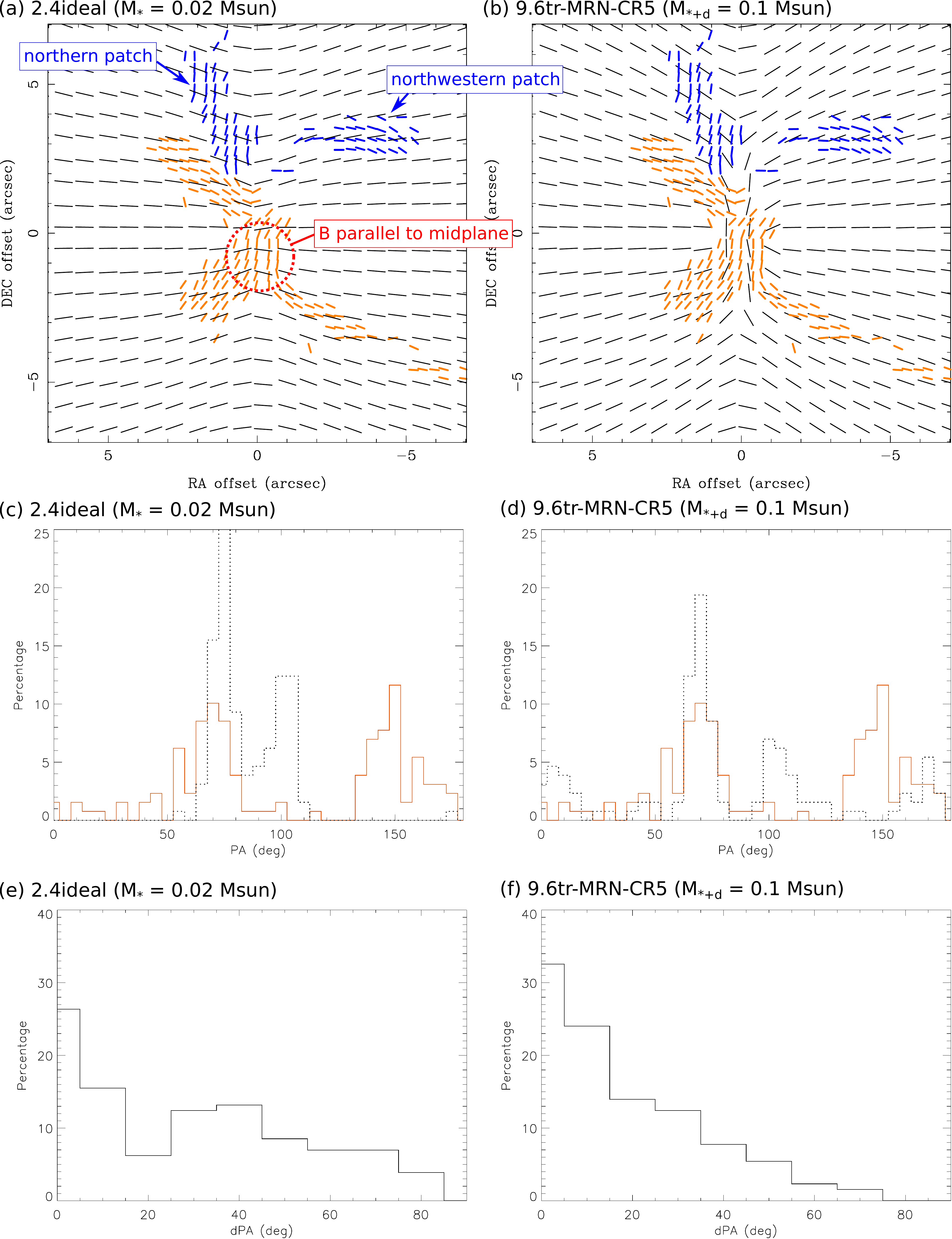}
\caption{(a) \& (b) Comparison of the magnetic field orientations from the ALMA observations (blue and orange segments) and from the synthetic maps (black segments) of our models, 2.4Ideal with the stellar mass of 0.02 $M_\odot$ (left) and 9.6tr-MRN-CR5 with the star+disk mass of 0.1 $M_\odot$ (right), respectively. The center of the map, (0, 0), is the protostellar position. (c) and (d) Number fraction distributions of the magnetic field orientations from the ALMA observations (orange histogram) and the models (black dotted histograms). Only the orange segments in (a) and (b) and the model segments at the same positions are included in these plots. (e) \& (f) Number fraction distributions of the PA difference in the magnetic field orientations from the ALMA observations and the models. The northern and northwestern patches discussed in Section \ref{obvssim} and \ref{discuss} are labeled in (a). A red circle in (a) denotes the region showing the magnetic field orientation almost parallel to the midplane.}\label{simpol2}
\end{figure*}

Figure~\ref{simpol2}a and b present the comparison between the magnetic field orientations on a 1000 au scale from the ALMA observations and from our models 9.6tr-MRN-CR5 with $M_{\star{\rm+disk}}$ of 0.1 $M_\sun$ and 2.4Ideal with $M_\star$ of 0.02 $M_\sun$, which are the models showing the most and least pinched field among our models, respectively. 
The overall morphology of the observed magnetic field orientations is similar to the pinched magnetic field in our MHD simulations, except for the detections in the northern and northwestern patches (shown as blue segments). 
The northern patch of the magnetic field segments is oriented along the north--south direction (PA of 0\degr), which is the direction of the midplane. 
Although our model 9.6tr-MRN-CR5 also shows the magnetic field segments along the direction of the midplane, 
these segments are located close to the midplane and within a radius of 2$\arcsec$ (200 au).
All the segments  at a radius larger than 200 au are tilted from the direction of the midplane in our models.
On the other hand, the orientations of the northwestern magnetic field segments change from the northwest--southeast direction with a PA of $\sim$120$\degr$ to the northeast--southwest direction with a PA of $\sim$50$\degr$ as the distance to the center increases. 
At the outer radii larger than 250 au (RA offsets $<-2\farcs5$), the observed orientations become perpendicular to those in our model maps.

\begin{figure}
\centering
\includegraphics[width=8cm]{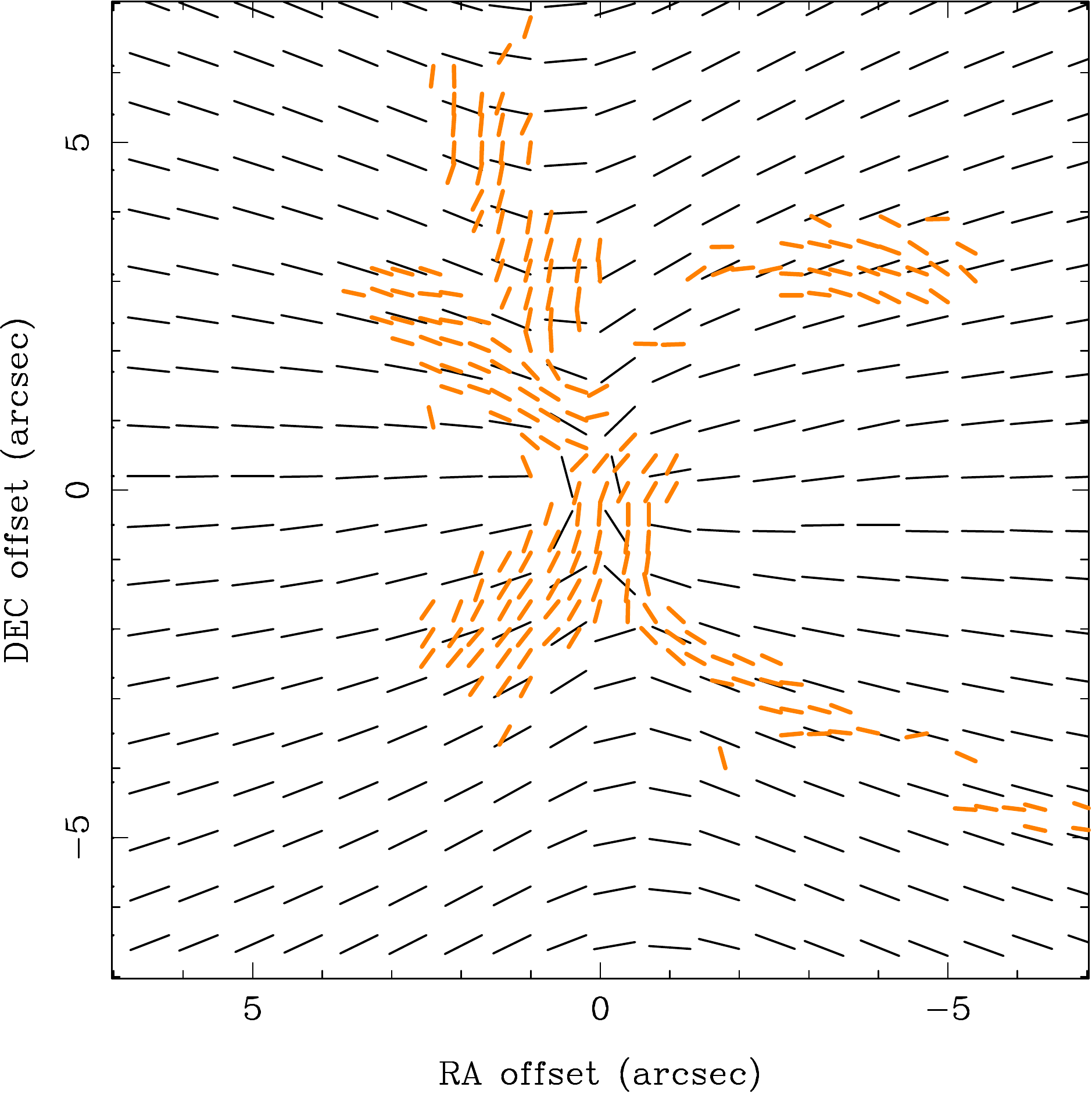}
\caption{Same as Fig.~\ref{simpol2} (a) and (b) but for model mis-tr-MRN-CR5, where the magnetic field and rotation axis are misaligned by 15$
\degr$.}
\label{simpol3}
\end{figure}

We excluded the northern and northwestern patches of the detections and compared the number distributions of the magnetic field orientations from the observations (orange segments) with the models in Fig.~\ref{simpol2}c and d.
The northeastern and southwestern segments are oriented along the northeast--southwest direction and appear as a peak at PA of 70$\degr$ in the number distribution. 
These segments can be well explained with the pinched magnetic field in our models. 
The segments observed around the center are mostly oriented close to the direction of the midplane. 
Such an orientation of the magnetic field segments around the center can also be explained with our weak field models with initial $\lambda$ of 9.6 at the later evolutionary time when $M_{\rm \star+disk}$ is 0.1 $M_\sun$.
We found that all of our models with the stronger field of initial $\lambda$ of 2.4 and 4.8 do not show such magnetic field segments along the midplane direction even at the later evolutionary time. 
In these models, the magnetic field segments around the center are more perpendicular to the midplane, as the case of 2.4Ideal shown in Fig.~\ref{simpol2}c. 
The observed magnetic field segments in the southeast are oriented along the northwest--southeast direction, corresponding to the peak at PA close to 150$\degr$ in the number distribution. 
Although the overall orientation of these segments is similar to those in our models, 
the segments in our models are more tilted away from the midplane and have a PA of 100$\degr$, as seen in the number distribution (Fig.~\ref{simpol2}d). 
The magnetic field segments with PA of $\sim$150$\degr$ are also present in the southeast in our model map of the weak field case with initial $\lambda$ of 9.6 (Fig.~\ref{simpol2}b), 
but they are located closer to the midplane and do not extended to 1$\arcsec$--2$\arcsec$ from the midplane as the observations. 
Additionally, we also performed simulations with the magnetic field and rotation axis misaligned by 15$\degr$. 
We rotated the magnetic field direction to have a PA of 90$\degr$, when we generated the synthetic maps of these simulations with the misalignment. 
Although the model maps from the simulations with the misalignment show different PAs of the magnetic field orientations in the northwest and the southeast, compared to those with the aligned magnetic field and rotational axis (Fig.~\ref{simpol3}), 
incorporating the misalignment still cannot explain the observed magnetic field orientations in the northwest and the southeast. 

In summary, 
there are two patches of the magnetic field orientations observed in the north and the northwest that cannot be explained with our MHD simulations. 
The possible reasons are discussed in Section \ref{discuss}. 
The other observed magnetic field orientations can be explained with the pinched magnetic field in our simulations, 
and we find that the simulations with the weak field of initial $\lambda$ of 9.6 explain the observations better than our other simulations with the stronger field (e.g., Fig.~\ref{simpol2}e and f).
Especially, the presence of the magnetic field orientations parallel to the midplane observed around the center can only be explained with the models with initial $\lambda$ of 9.6 at the later evolutionary time when $M_{\rm \star+disk}$ is 0.1 $M_\sun$ among all of our models.
This is similar to the results by \citet{Maury18}. 
They found that the observed magnetic field orientations are similar to their models with the largest mass-to-flux ratio\footnote{Because of the different definitions of the mass-to-flux ratio, the mass-to-flux ratio $\mu$ in \citet{Maury18} is ${3 \over \sqrt{10}}\lambda$.} $\mu$ of 6, corresponding to $\lambda$ of 6.3.
Nevertheless, the observed magnetic field segments in the southeast are oriented closer to the midplane by $\sim$30$\degr$ compared to those in our models. 
In addition, we note that the observed magnetic field structures are not symmetric with respect to the magnetic field direction of PA of $\sim$90$\degr$. 
The number distribution of the observed magnetic field orientations has two peaks at PA of 70$\degr$ and 150$\degr$ (Fig.~\ref{simpol2}), 
and thus, the symmetric axis is at PA of 110$\degr$. 
This distribution is different from that in our models having the peaks at PA of 70$\degr$ and 105$\degr$ with the magnetic field direction of PA of 90$\degr$. 

\section{Discussions}\label{discuss}

\begin{figure*}
\centering
\includegraphics[width=16cm]{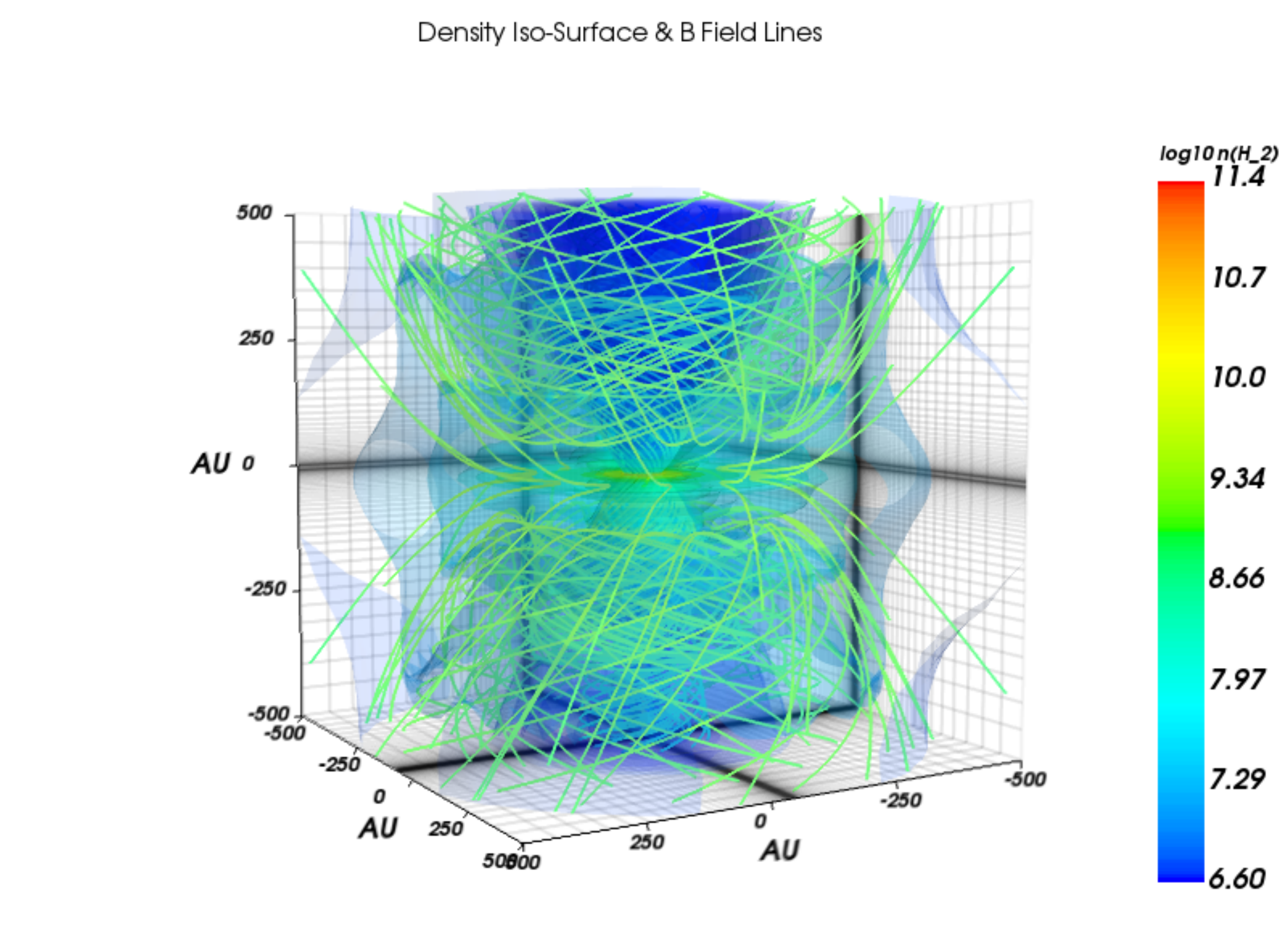}
\caption{Three-dimensional structures of the magnetic field lines in our model 9.6tr-MRN-CR5 (green lines). Color scale presents the density.}\label{3dfield}
\end{figure*}

\subsection{Discrepancy between observations and simulations}\label{discuss1}
Although the overall morphology of the observed magnetic field in B335 is similar to the pinched field in our MHD simulations (Fig.~\ref{simpol1}), 
the symmetric axis of the magnetic field orientations on a 1000 au scale observed with ALMA is likely at PA of $\sim$110$\degr$, different from the magnetic field direction of 87$\degr$--94$\degr$ on a 6000 au scale inferred from the JCMT observations. 
In addition, a part of the magnetic field segments in the northwest observed with ALMA are almost perpendicular to those in our simulations (Fig.~\ref{simpol2}). 
We found that in our MHD simulations, the magnetic field is highly wrapped around the bipolar outflow, 
and there are significant toroidal components of the magnetic field in the wall of the outflow cavity (Fig.~\ref{3dfield}). 
Because the outflow in our simulations are more or less axisymmetric, these toroidal components are cancelled out after the integration along the line of sight in our radiative transfer calculation, 
and only the poloidal components that are aligned with or moderately tilted from the axis of the bipolar outflow are seen in the region of the outflow in our synthetic maps (Fig.~\ref{simpol2}). 
In the ALMA observations, the polarized emission is primarily detected in the wall of the outflow cavity.  
The structures of the bipolar outflow in B335 are clearly asymmetric as observed in the CO line \citep{Yen10} and in the continuum \citep[Fig.~\ref{bmap}b][]{Maury18}. 
Thus, the inhomogeneous density structures (if any) in the wall of the outflow cavity could affect the observed polarization orientations in B335 because the magnetic field direction could change a lot along the line of sight passing through the outflow, as seen in our MHD simulations. 
As presented in Fig.~\ref{3dfield}, in our simulations, there are indeed magnetic field lines perpendicular to the wall of the outflow cavity. 
This could explain the observed magnetic field segments in the northwestern region in B335 with ALMA.
Therefore, the discrepancy in the magnetic field orientations in the northwest in the observations and the simulations could be due to the effect of the integration along the line of sight passing through the asymmetric outflow. 

The northern patch of the magnetic field orientations observed with ALMA, which are almost parallel to the midplane, cannot be reproduced in our MHD simulations regardless of the field strengths considered. 
As suggested by our JCMT results, 
the magnetic field direction on the large scale is most likely perpendicular to the midplace. 
To form magnetic field lines severely pinched and parallel to the midplane on a 1000 au scale as observed with ALMA, 
one possibility is that there is an accretion flow with a high mass infalling rate, a weak field strength, and low magnetic diffusion from the northern region in B335, 
as discussed in Section \ref{pinch}.
The severely pinched magnetic field formed by a strong accretion flow in B335 was also suggested by \citet{Maury18}. 
Nevertheless, such magnetic field orientations parallel to the midplane are not detected in the southern region. 
If the magnetic field orientations parallel to the midplane in the north are indeed formed by a strong accretion flow, 
the results could hint that the distribution of the collapsing material is not symmetric. 
An asymmetric accretion flow dragging the magnetic field line to form a severely pinched magnetic field at a radius of 200--500 au is not present in our initially axisymmetric MHD simulations or our simulations with the misaligned magnetic field and rotational axis. 
Asymmetric accretion flows have been seen in other MHD simulations with turbulence, such as those in \citet{Li14}, where the pseudo disk and thus the accretion flow are warped because of the turbulence. 
On the other hand, we note that there is no polarized emission detected in the northeastern and southern regions with our JCMT observations. 
This non-detection could hint at the presence of complex magnetic field structures in those regions. 
Consequencely, the polarized emission with different polarization orientations could be cancelled out. 
ALMA polarimetric observations with a wider field of view and a high resolution are needed to image complete field structures in B335. 

\subsection{Magnetic field strength and gas kinematics in B335}
The magnetic field structures on a 6000 au scale observed with the JCMT can be well explained with the pinched field in our MHD simulations, 
but the angular resolution of the JCMT observations is insufficient to distinguish different degrees of the pinched field in our simulations with different field strengths (Fig.~\ref{simpol1}).
The comparison between our MHD simulations and the ALMA observations shows that the observed magnetic field structures on a 1000 au scale are better explained with our simulations with the weak magnetic field of initial $\lambda$ of 9.6 than with the strong magnetic field of initial $\lambda$ of 2.4 and 4.8 (Fig.~\ref{simpol2}).
Especially, the observed magnetic field orientations close to the center within a radius of 200 au are almost parallel to the midplane. 
Such field orientations are only seen in our simulations with the weak field at the later evolutionary time when $M_{\star+\rm disk}$ is 0.1 $M_\sun$. 
Thus, these results favor a weak magnetic field of initial $\lambda$ of 9.6 in B335, as also suggested by \citet{Maury18}.

However, we also found that with the weak magnetic field, a Keplerian disk with a radius of 10 au quickly forms and grows to have a radius of 50 au when $M_{\star+\rm disk}$ is 0.1 $M_\sun$ in our tr-MRN simulations with a typical cosmic-ray ionzation rate of 10$^{-17}$ s$^{-1}$, while no disk forms in our ideal MHD or MRN simulations.
The ALMA observations in the C$^{18}$O and SO lines at an angular resolution of 0\farcs3 (30 au) did not find any sign of Keplerian rotation and put an upper limit of the disk radius of 10 au in B335 \citep{Yen15b}. 
On the other hand, the fan-like bipolar outflows are observed on scales from hundreds to thousands of au in B335 \citep{Hir88, Yen10, Hull14}. 
Hence, a small rotationally supported disk is likely present around the protostar to launch such outflows via a magneto-centrifugal mechanism \citep[e.g.,][]{Blandford82, Zhao18a}. 
Therefore, the presence of a small Keplerian disk in B335 favors the tr-MRN over ideal MHD and MRN simulations. 
Nevertheless, the radius of the disk formed in the tr-MRN simulation with the weak magnetic field and the typical cosmic-ray ionzation rate is larger than the observational upper limit. 
To reconcile the weak magnetic field and the small disk in B335, 
we find that the cosmic-ray ionzation rate needs to be higher than the typical value by a factor of five or more in order to strengthen the coupling between the magnetic field and matter which then increases the efficiency of magnetic braking. 
Our tr-MRN simulation with the weak magnetic field and a higher cosmic-ray ionzation rate of $5 \times 10^{-17}$ s$^{-1}$ forms a small Keplerian disk with a radius smaller than 10 au, 
and can explain the observed magnetic field parallel to the midplane within a radius of 200 au (Fig.~\ref{simpol1}b).

\begin{figure}
\centering
\includegraphics[width=8cm]{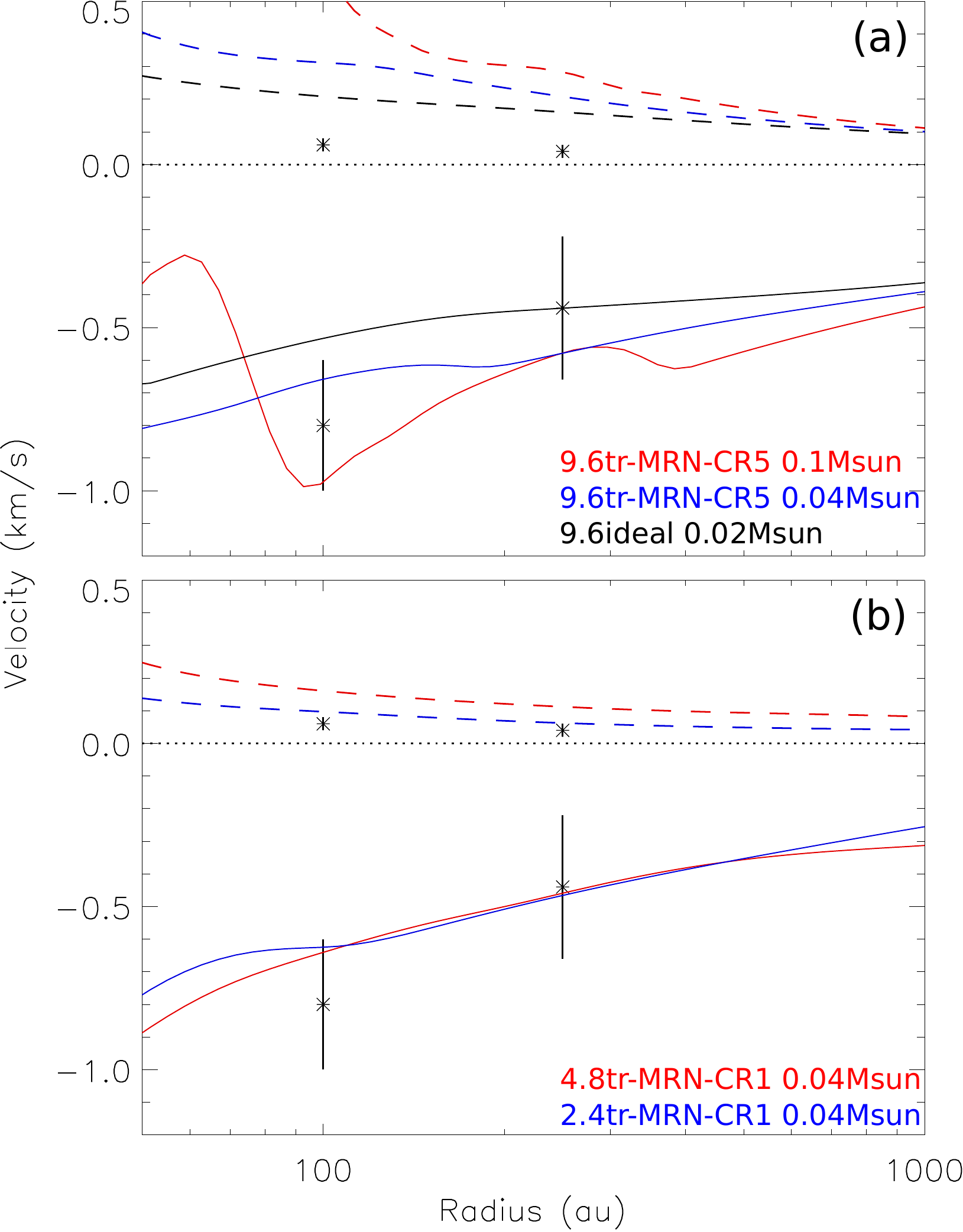}
\caption{Radial profiles of infalling (solid lines) and rotational (dashed lines) velocities along the midplane from the models with (a) the weak magnetic field of initial $\lambda$ of 9.6 and (b) the stronger magnetic field of initial $\lambda$ of 2.4 and 4.8 in comparison with the observational measurements (data points) obtained with the SMA and ALMA. Different models are denoted with different colors and are labeled at the bottom right corner in each panel. Because B335 is almost edge on, the uncertainty of the estimated rotational velocity, which is 10\%--20\%, is smaller than that of the estimated infalling velocity \citep{Yen15b}. Thus, the error bars of the rotational velocity cannot be seen clearly in the figures.}
\label{simvr}
\end{figure}

In Fig.~\ref{simvr}, we compare the radial profiles of the infalling and rotational velocities along the midplane in the simulations and the observational estimates. 
The infalling velocities on a scale of 100--250 au in the simulations with different magnetic field strengths are all consistent with the observations within the uncertainties because in these simulations, $M_{\rm \star+disk}$ is in the range of the observational estimates. 
In contrast to that, 
the rotational velocities in the simulations with the weak magnetic field and the high cosmic-ray ionization rate are higher than those estimated from the observations by a factor of four to ten, suggesting that with the weak magnetic field of initial $\lambda$ of 9.6, magnetic braking is not efficient to slow down the rotational motion. 
As also discussed in \citet{Yen15b}, 
the observed specific angular momentum of the collapsing material on a 100 au scale in B335 is a factor of two lower than the expectation from the inside-out collapse model where the angular momentum is conserved, suggestive of efficient magnetic braking in the inner envelope. 
We note that even in our ideal MHD simulations, the rotational velocity in the simulations is still higher than in the observations by a factor of three, when the initial magnetic field strength is $\lambda$ of 9.6.
Thus, the higher rotational velocity and inefficient magnetic braking in these simulations are due to the magnetic field strength but not the non-ideal MHD effect. 
Only when the initial magnetic field strength is stronger than $\lambda$ of 4.8, the rotational velocities in the simulations and the observations become comparable within a factor of two.
Therefore, the comparison of the rotational velocities in the simulations and the observations could favor the strong magnetic field of initial $\lambda < 4.8$ in B335. 
This is different from the inferred magnetic field strength of $\lambda$ of 9.6 based on the polarimetric observations described above. 

There are two possibilities resulting in the different magnetic field strengths inferred from the polarimetric and molecular-line observations: (1) the rotational-to-gravitational energy $\beta_{\rm rot}$ is overestimated and (2) additional contributions in the polarized intensity from other mechanisms, such as dust scattering.
In our MHD simulations, $\beta_{\rm rot}$ is adopted to be 0.4\% based on the observational estimates of the core mass of $\sim$1 $M_\sun$ and the angular speed of the core rotation of $4 \times 10^{-14}$ s$^{-1}$.
The angular speed was estimated based on the global velocity gradient along the major axis of the dense core observed with the single-dish telescopes \citep{Saito99, Yen11, Kurono13}. 
Numerical simulations of dense cores including synthetic observations show that the specific angular momentum derived from the synthetic images of the dense cores can be a factor of 8--10 higher than their actual specific angular momentum computed by the sum of the angular momenta contributed by the individual gas parcels in the dense cores \citep{Dib10}.  
In addition, if there are filamentary structures in the dense core in B335, which could not be resolved with the single-dish observations, 
infalling motions along the filamentary structures could also contribute to the observed velocity gradient, leading to an overestimated angular speed of the core rotation \citep{Tobin12}.
We have also performed our simulations with a lower $\beta_{\rm rot}$, 
and we find that the rotational velocity on a 100 au scale in the simulations decreases with decreasing $\beta_{\rm rot}$.
Thus, the discrepancy in the magnetic field strengths inferred from the field structures and the gas kinematics can be reconciled, if the core rotation in B335 is overestimated by a factor of a few in the observations, 
and these results would suggest a weak magnetic field of initial $\lambda$ of 9.6 in B335.
Further observations combining single dishes and interferometers to have a high spatial dynamical range and to map the velocity structures of the entire dense core in B335 at a high angular resolution are needed to study coherent velocity features and provide a better estimate of the core rotation. 

The other possibility is that the polarized emission observed close to the center could have contributions from dust scattering. 
If large dust grains with a size of the order of 100 $\mu$m are present in the inner dense region on a 100 au scale, where the density is higher than 10$^7$ cm$^{-3}$ \citep{Harvey03, Evans15}, scattering of anisotropic continuum emission by these dust grains could induce polarized emission with a polarization percentage of $\sim$1\% \citep{Kataoka15, Yang16, Yang17}, similar to that observed around the center with ALMA \citep{Maury18}. 
In addition, the polarization orientations close to the midplane around the center observed with ALMA are almost perpendicular to the midplane, 
and thus, the inferred magnetic field orientations are parallel to the midplane \citep{Maury18}, 
as shown in Fig.~\ref{bmap}.
The presence of the polarization orientations perpendicular to the midplane are consistent with the expectation of scattering-induced polarization in an edge-on disk \citep{Yang16, Yang17}, 
and B335 is indeed an edge-on source \citep{Hir88, Stutz08}.
Signatures of scattering-induced polarized emission have been observed in a few embedded young protostars \citep{Stephens17, Cox18, Lee18}.
Thus, because of the contribution of the scattering, 
the actual magnetic field structures close to center may not be as severely pinched as discussed in Section \ref{smallB}. 
In this case, these results would favor the strong magnetic field of initial $\lambda < 4.8$ in B335. 
Nevertheless, this scenario of the scattering-induced polarization cannot explain the inferred magnetic field orientations almost parallel to the midplane at a radius of $>$200 au in the north (Section \ref{discuss1}), where the grains are unlikely large enough to induce any significant polarzation through scattering. 
Future ALMA polarimetric observations at higher angular resolutions and at different wavelengths are required to resolve the polarization emission on a 100 au scale in B335 and to study its nature.

\section{Summary}
We present the results and analysis of our JCMT POL-2 observations at 850 $\mu$m and the ALMA archival polarimetric data at 1.3 mm of B335. 
In addition, we carried out a series of (non-)ideal MHD simulations of the collapse of a rotating non-turbulent dense core, whose mass and angular speed of the rotation are adopted to be the same as those observed in B335. We generated synthetic polarization maps from these simulations to compare with the polarimetric data. 
Our main results are summarized below. 

\begin{enumerate}
\item{We carried out MHD simulations with different magnetic field strengths of $\lambda$ of 2.4, 4.8, and 9.6, cosmic-ray ionization rates of 1, 5, and 10 $
\times$ 10$^{-17}$ s$^{-1}$, and distributions of dust grain sizes with minimum sizes of 0.005 $\mu$m and 0.1 $\mu$m. 
The latter two parameters determine the magnetic diffusivity.
With the initial core rotation and mass set to be the same as in B335, 
we find that a persistent disk only forms in the simulations with the minimum grain size of 0.1 $\mu$m, 
where ambipolar diffusion is enhanced because of the removal of the small grains, regardless of other parameters. 
In our simulations, the magnetic field lines generally show an hour-glass morphology. 
We find that the degree of the pinch of the magnetic field is regulated by the Lorentz force, 
and that the magnetic field in the simulations with weaker field strength, lower magnetic diffusivity, and longer evolutionary time is pinched more severely. 
}
\item{The magnetic field orientations on a 6000 au scale in B335 inferred from the JCMT POL-2 observations are along the east--west direction, well aligned with the direction of the outflow within 10$\degr$ on the plane of the sky.
The observed magnetic field structures with the JCMT can be well explained with all our simulations with different magnetic field strengths. 
The comparison between our JCMT and simulation results suggest that the magnetic field on a 6000 au scale in B335 is pinched. 
The JCMT resolution is not sufficient to distinguish different magnetic field structures in our simulations with different field strengths.
}
\item{The ALMA polarization results show the signatures of a pinched magnetic field on a 1000 au scale in B335, which are similar to our MHD simulations. 
In addition, the magnetic field orientations close to the center and at radii of 200--500 au in the north are almost parallel to the midplane. 
The observed field orientations in the north cannot be explained with our simulations, where the magnetic field orientations are tilted from the midplane. 
These magnetic field structures could be caused by a strong accretion flow from the north. 
Among all our MHD simulations, the observed magnetic field orientations almost parallel to the midplane in the central region can only be explained with the weak field models having an initial mass-to-flux ratio of 9.6 at the late evolutionary time when the mass of the central star+disk system is 0.1~$M_\sun$.
}
\item{The comparison between the ALMA and simulation results favor weak field models with an initial mass-to-flux ratio of 9.6. 
Nevertheless, with such a weak magnetic field, 
the rotational velocity on a 100 au scale and the size of the Keplerian disk formed in our simulations are a factor of several higher than the observational estimates. 
We find that when the cosmic-ray ionization rate is increased by a factor of five or more to enhance the field--matter coupling and thus the efficiency of magnetic braking, 
the disk size in our weak field simulations is reduced and consistent with the observational estimate,
while the rotational velocity on a 100 scale in the simulations is still higher than in the observations. 
On the other hand, 
the rotational velocity on a 100 au scale and the disk size in the simulations both become comparable to the observations in our stronger field models with an initial mass-to-flux ratio smaller than 4.8, regardless of magnetic diffusivity. 
Thus, the comparison between the observed gas kinematics in B335 and our MHD simulations favors the stronger field models.
}
\item{There are two possibilities resulting in the different magnetic field strengths inferred from the magnetic field structures and the gas kinematics: (1) an overestimated rotational-to-gravitational energy $\beta_{\rm rot}$ and (2) additional contributions in the polarized intensity from dust scattering. 
The rotational velocity on a 100 au scale in the simulations is proportional to $\beta_{\rm rot}$ of the initial core. 
The presence of incoherent gas motions or infalling motion along filamentary structures (if any) in the dense core in B335 could lead to an overestimated $\beta_{\rm rot}$ from the global velocity gradient observed with single-dish telescopes with limited resolutions. 
In this case, our results favor the weak field models for B335. 
On the other hand, dust scattering in an edge-on source like B335 tends to induce a polarization orientation perpendicular to the midplane. 
This effect could make the inferred magnetic field orientations in the central dense region more parallel to the midplane, 
and the actual magnetic field structures might not be severely pinched. 
In this case, based on the observed gas kinematics, our results favor the strong field models for B335.
}
\end{enumerate}

\begin{acknowledgements} 
This paper makes use of the following ALMA data: ADS/JAO.ALMA\#2013.1.01380.S. ALMA is a partnership of ESO (representing its member states), NSF (USA) and NINS (Japan), together with NRC (Canada), MOST and ASIAA (Taiwan), and KASI (Republic of Korea), in cooperation with the Republic of Chile. The Joint ALMA Observatory is operated by ESO, AUI/NRAO and NAOJ. 
The JCMT data were obtained under program ID M17AP067. 
JCMT is operated by the East Asian Observatory on behalf of The National Astronomical Observatory of Japan; Academia Sinica Institute of Astronomy and Astrophysics; the Korea Astronomy and Space Science Institute; the Operation, Maintenance and Upgrading Fund for Astronomical Telescopes and Facility Instruments, budgeted from the Ministry of Finance (MOF) of China and administrated by the Chinese Academy of Sciences (CAS), as well as the National Key R\&D Program of China (No. 2017YFA0402700). Additional funding support is provided by the Science and Technology Facilities Council of the United Kingdom and participating universities in the United Kingdom and Canada.
We thank all the ALMA and JCMT staff supporting this work. 
P.M.K. acknowledges support from MOST 107-2119-M-001-023 and from an Academia Sinica Career Development Award. 
ZYL is supported in part by NSF AST-1716259 and 1815784 and NASA 80NSSC18K1095 and NNX14AB38G. 
S.T. acknowledges a grant from JSPS KAKENHI Grant Numbers JP16H07086 and JP18K03703 in support of this work. 
This work was supported by NAOJ ALMA Scientific Research Grant Numbers 2017-04A. 
\end{acknowledgements} 

\begin{appendix}
\section{Simulated and convolved model maps}\label{simpol}
\begin{figure}
\centering
\includegraphics[width=16cm]{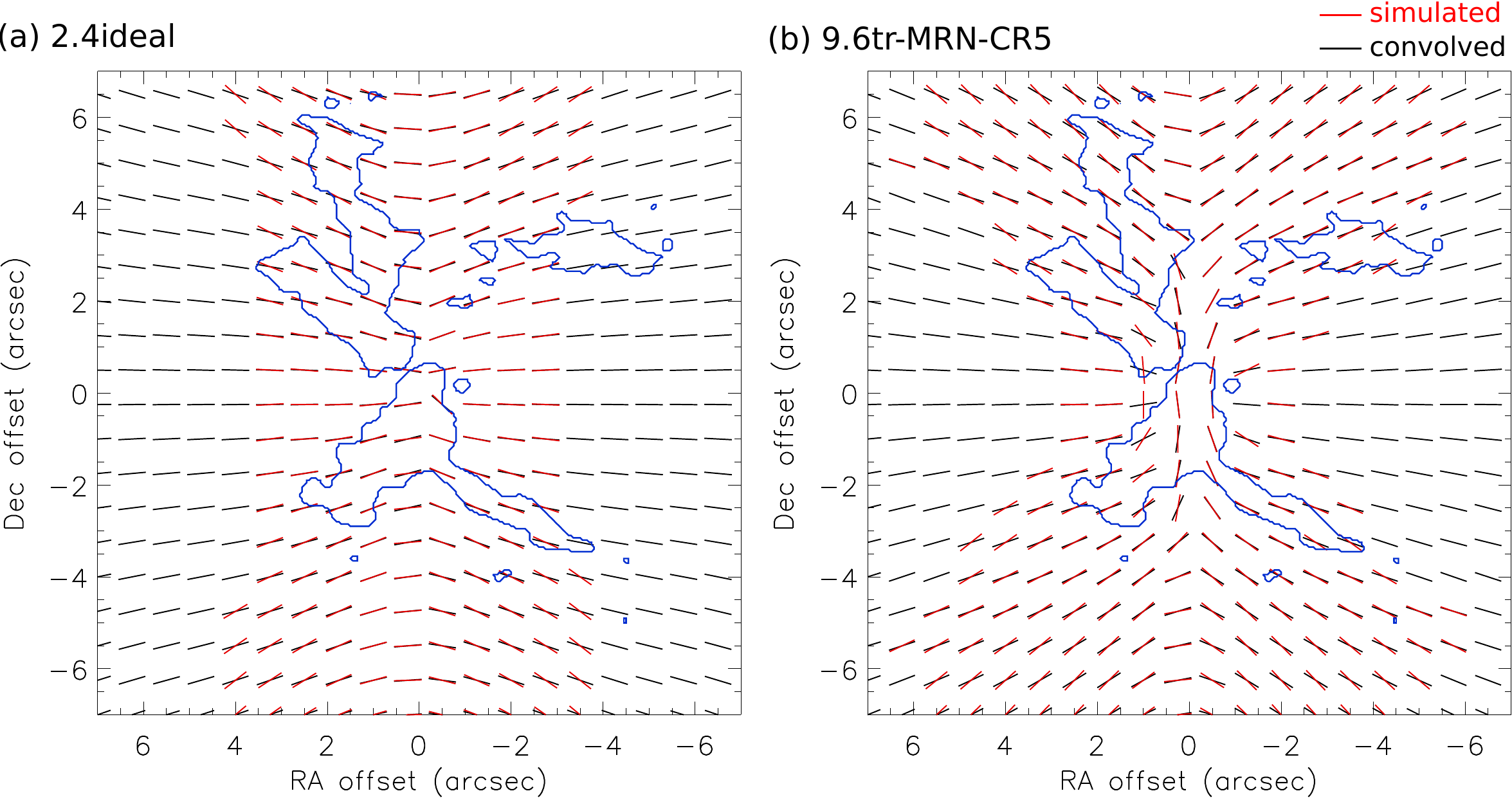}
\caption{Comparison between the magnetic field orientations extracted from the convolved (black segments) and simulated (red segments) model maps for the two cases with the least and most pinched magnetic field, (a) 2.4Ideal and (b) 9.6tr-MRN-CR5. Blue contours delineate the region where there are ALMA polarization detections. Because of the filtering effect of interferometry, smooth extended structures are filtered out in the simulated model maps, and thus there is no segment in the outer region along the outflow axis extracted from the simulated model maps.}
\label{modelpol}
\end{figure}

We performed imaging simulations of the ALMA observations on the synthetic Stokes {\it IQU} maps from two models, 2.4Ideal and 9.6tr-MRN-CR5, using the CASA simulator. 
These two models show the least and most pinched magnetic field among our models, respectively. 
The same array configuration as that in the ALMA observations was adopted in the imaging simulations.
The synthesized beam in our simulated maps is 0\farcs83 $\times$ 0\farcs47 comparable to that in the observations.
We extracted the polarization orientations from the simulated Stokes {\it Q} and {\it U} maps and rotated them by 90$\arcdeg$ to compare with those from the convolved model maps.
Figure \ref{modelpol} compares the magnetic field orientations extracted from the simulated and convolved model maps.
The numbers of the magnetic field orientations extracted from the simulated model maps are smaller than those from the convolved model maps because the extended emission in the outer region along the outflow axis is filtered out in the simulated observations.
For model 2.4Ideal, there is no significant difference in the magnetic field orientations extracted from the simulated and convolved model maps.
For model 9.6tr-MRN-CR5, the magnetic field orientations extracted from the simulated model map are generally consistent with those from the convolved model maps. 
Few segments at offsets close to (1$\arcsec$, 0\arcsec) extracted from the simulated model maps are orientated more along the direction of the mid-plane, 
while those from the convolved model maps are more perpendicular to the mid-plane.
These segments are located near the boundary between the outflow and the inner flatten envelope, where the directions of the magnetic field lines change significantly (e.g., Fig.~\ref{comp_mhds}). 
Nevertheless, in the region where there are ALMA polarization detections, there is no significant difference between the magnetic field orientations extracted from the simulated and convolved model maps for model 9.6tr-MRN-CR5. 
Therefore, our comparison between the models and the observations is not affected by the filtering effect of the ALMA observations.

\end{appendix}

\end{document}